\documentclass[%
reprint,
superscriptaddress,
 amsmath,amssymb,
prm,
]{revtex4-2}
\usepackage{comment}
\usepackage{graphicx}
\usepackage{dcolumn}
\usepackage{bm}
\usepackage{hyperref}
\usepackage{subfig}
\usepackage{xcolor}

\begin{document}

\newcommand{\blue}[1]{\textcolor{black}{#1}}

\preprint{APS/123-QED}

\title{A collinear-spin machine learned interatomic potential for Fe\textsubscript{7}Cr\textsubscript{2}Ni alloy}

\author{Lakshmi Shenoy}
\email[]{lakshmi.shenoy@warwick.ac.uk}
\affiliation{Warwick Centre for Predictive Modelling, School of Engineering, University of Warwick, Coventry CV4 7AL, United Kingdom}
\author{Christopher D. Woodgate}
\affiliation{Department of Physics, University of Warwick, Coventry CV4 7AL, United Kingdom}
\author{Julie B. Staunton}
\affiliation{Department of Physics, University of Warwick, Coventry CV4 7AL, United Kingdom}
\author{Albert P. Bart\'ok}
\affiliation{Warwick Centre for Predictive Modelling, School of Engineering, University of Warwick, Coventry CV4 7AL, United Kingdom}
\affiliation{Department of Physics, University of Warwick, Coventry CV4 7AL, United Kingdom}
\author{Charlotte S. Becquart}
\affiliation{Universite des Sciences et Technologies de Lille, Batiment C6, F-59655 Villeneuve d’Ascq Cedex, France}
\author{Christophe Domain}
\affiliation{Electricite de France, EDF Recherche et Developpement, Departement Materiaux et Mecanique des Composants, Les Renardieres, F-77250 Moret sur Loing, France}
\author{James R. Kermode}
\affiliation{Warwick Centre for Predictive Modelling, School of Engineering, University of Warwick, Coventry CV4 7AL, United Kingdom}

\date{\today}

\begin{abstract}

We have developed a new machine learned interatomic potential for the prototypical austenitic steel Fe$_{7}$Cr$_{2}$Ni, using the Gaussian approximation potential (GAP) framework. This new GAP can model the alloy's properties with \blue{close to density functional theory (DFT) accuracy, while at the same time allowing us to access larger length and time scales than expensive first-principles methods.} We also extended the GAP input descriptors to approximate the effects of collinear spins (Spin GAP), and demonstrate how this extended model \blue{successfully predicts structural distortions  due to antiferromagnetic and paramagnetic spin states}. We demonstrate the application of the Spin GAP model for bulk properties and vacancies and validate against DFT. These results are a step towards modelling \blue{ the atomistic origins of ageing in austenitic steels with higher accuracy.}

\end{abstract}

\maketitle


\section{\label{sec:intro}
Introduction}

Austenitic stainless steels are key structural materials with high mechanical strength and corrosion resistance. They have many applications, from everyday tools such as household appliances and cars, to structural components of buildings, bridges and industrial machines \cite{parr1993stainless}. Their desirable properties come from alloying Fe with a concentration of 16-20\% Cr for rust resistance, and about 10\% Ni to maintain a crack-resistant face-centred-cubic crystal structure at all temperatures where it remains solid ($<$~1800~K) \cite{marshall_austenitic_1984}. Specific properties of the steel can be fine tuned by adding small concentrations ($<$~1\%) of other solutes such as Mo, Mn, P, C etc, depending on the application of interest \cite{ledbetter_effects_1985}. However the core properties come from the Fe-Cr-Ni mix, so in this work we focus on the prototypical austenitic steel with composition Fe$_{7}$Cr$_{2}$Ni.

A key application  of austenitic steels is in nuclear power plants. The austenitic steel grades 304 and 316, which have a composition very close to the model alloy Fe$_{7}$Cr$_{2}$Ni, are used in the primary containment barriers of light-water nuclear reactor pressure vessels (RPV) \cite{was_fundamentals_2007}. The RPV interiors are exposed to constant neutron bombardment and a wide range of thermal fluctuations from cryogenic temperatures during maintenance to operating temperature of $\sim$~600~K. Over time, these harsh conditions trigger the formation of radiation-induced point defects in the RPV interiors, which eventually cluster to form larger defects such as voids and dislocation loops. Macroscopically, these manifest as creep, swelling, stress corrosion, and other ageing effects which are safety concerns \cite{piochaud_first-principles_2014}. To improve RPV designs and facilitate mitigation efforts, there is interest in understanding the mechanisms behind these ageing phenomena.

Experimental data for aging phenomena are limited due to the challenges in conducting controlled experiments in extreme conditions, and also challenges in measuring atomic quantities experimentally \cite{dimitrov_defect_1982,dimitrov_composition_1984,benkaddour_determination_2021}. We therefore require computational methods to simulate these phenomena across length scales, to fill in the gaps between the available experimental data, and to explain the observed aging effects. 

\blue{There are various tiers of first principles methods available for materials modelling, with higher accuracy being accompanied by (prohibitively) higher computational costs. While more complex phenomena such as nuclear cascades or spectroscopic excitations require higher levels of theory, the first principles method of choice for modelling mechanical properties as in our case} is density functional theory (DFT), in which the electronic structure of materials is calculated from approximations of quantum mechanics \cite{kohn_self-consistent_1965}. There have been efforts to study the Fe$_{7}$Cr$_{2}$Ni alloy using DFT \cite{piochaud_first-principles_2014,manzoor_factors_2021,zhang_statistical_2021,antillon2022b}. However, as the computational cost of DFT calculations scales cubically with system size, these studies were restricted to supercells containing upto 256 atoms. These supercells are too small to model extended defects such as voids or dislocations. Also, as DFT calculations are computationally expensive, these studies were restricted to limited statistics, which is not ideal for modelling the variations due to chemical complexity of concentrated alloys.

Computationally more affordable alternatives for atomistic modelling of materials that allow us to simulate larger length and time scales are classical interatomic potentials. These models have a fixed mathematical form and are fit to experimental and DFT data for the material properties of interest. For metallic systems, the main classical models used are the embedded atom models (EAM) \cite{johnson_alloy_1989}, the modified EAM \cite{jelinek_modified_2012}, and the Finnis Sinclair model \cite{rafii-tabar_long-range_1991}. While there have been many efforts to study alloys using classical interatomic potentials \cite{smith_application_1989,grujicic_analysis_1993,bonny_interatomic_2011,bonny_interatomic_2013,daramola_development_2022}, these studies are restricted to just a few properties owing to the fixed mathematical form of empirical models. For instance, Bonny \textit{et. al.} developed a FeCrNi EAM for plasticity \cite{bonny_interatomic_2011} and another FeCrNi EAM for point defects \cite{bonny_interatomic_2013} (referred to as EAM-13 throughout this paper), but neither can simulate both types of defects as they were trained for their respective target properties. We would ideally like to have a model that can simulate multiple defects and their interactions. 

A successful emerging approach that can address the shortcoming of both DFT and EAM approaches is machine learned interatomic potentials (MLIPs) \cite{deringer_machine_2019}. \blue{MLIPs are trained on a database of first-principles data, generally from DFT.} Their highly flexible mathematical forms allow them to interpolate the training data smoothly and so fit the potential energy surface with \blue{near DFT accuracy, while at the same time being much cheaper to use for predictions than first-principles methods directly.}  A major advantage of the MLIPs over classical potentials like EAM is that they can be trained on multiple defects, by including representative configurations in the training database. Moreover, the accuracy and scope of an MLIP can be improved iteratively by appending new training configurations to its training database, chosen either manually using knowledge about the system of interest or using active learning schemes \cite{van2022hyperactive}. 

A range of MLIP approaches have enjoyed considerable success, including kernel methods~\cite{bartok_gaussian_2010}
linear and non-linear expansions of the potential in a polynomial basis such as the atomic cluster expansion (ACE)~\cite{drautz_atomic_2019} or moment tensor potential (MTP) \cite{shapeev_moment_2016} frameworks,
and most recently message-passing neural networks \cite{batatia_mace_2022,batzner_e3-equivariant_2022,lot_panna_2020}.
In this study, we use the Gaussian approximation potential (GAP) framework, a kernel based method \cite{bartok_gaussian_2010}. The GAP model has physically motivated kernel choices and hyperparameters, making it intuitive to understand. It uses Gaussian process regression to fit the potential energy surface; the Bayesian interpretation of this procedure allows us to derive error estimates for GAP predictions, which is a useful feature of this choice of MLIP. The GAP model has been fitted and validated for many elemental materials such as Si \cite{bartok_machine_2018}, W \cite{szlachta_accuracy_2014}, Fe \cite{dragoni_achieving_2018} and C \cite{deringer_machine_2017}, demonstrating the effectiveness of this model. More recently, it has also been used for binary alloys \cite{rosenbrock_machine-learned_2021} and the ternary alloy Ge-Sb-Te \cite{mocanu_modeling_2018}. In this work, we fit and validate a GAP model for Fe$_{7}$Cr$_{2}$Ni. We also extend the GAP descriptors to incorporate the effects of the spin-polarised electronic structure of the alloy in terms of collinear spins associated with the Fe atoms, and demonstrate how this Spin GAP model improves predictions compared to the standard GAP model.  

This paper is organised as follows: Section \ref{sec:methods} outlines how the training database was assembled, and includes details of the standard GAP and Spin GAP fitting procedures. Section \ref{sec:results} discusses the main results for bulk and point defect properties of Fe$_{7}$Cr$_{2}$Ni using the Spin GAP. The results are validated against DFT and experiments, and compared with EAM and standard GAP predictions.

\section{\label{sec:methods}
Methodology}

\subsection{\label{subsec:database}
Training Database}
\begin{table}
    \centering
    \caption{\label{tab:database} \blue{Number of training and testing configurations for each subgroup DB\textit{x}, comprising either} 256-atom bulk or 255-atom vacancy configurations.}
    \begin{ruledtabular}
    \begin{tabular}{cccccc}
          & Config Type & Temperature & Train & \blue{Test}  \\ \hline
      DB1 & relaxed bulk & 0~K & 111 & \blue{12} \\
      DB2 & sheared bulk & $\sim$0~K & 190 & \blue{75} \\
      DB3 & relaxed vacancy & 0~K & 36 & \blue{8} \\
      DB4 & vacancy optimisation & $\sim$0~K & 72 & \blue{20} \\
      DB5 & bulk MD & 1000/1500/2000~K & 60 & \blue{13} \\
      DB6 & vacancy MD & 1000/1500/2000~K & 71 & \blue{20} \\
      DB7 & bulk MD quench & 1000~K $\rightarrow$ 0~K & 38 & \blue{4} \\
      DB8 & vacancy MD quench & 1000~K $\rightarrow$ 0~K & 44 & \blue{28} \\ 
       & \blue{element substitutions} & \blue{0K} & \blue{0} & \blue{42} \\
    \end{tabular}
    \end{ruledtabular}
\end{table}

\begin{figure}
    \includegraphics[width=\linewidth]{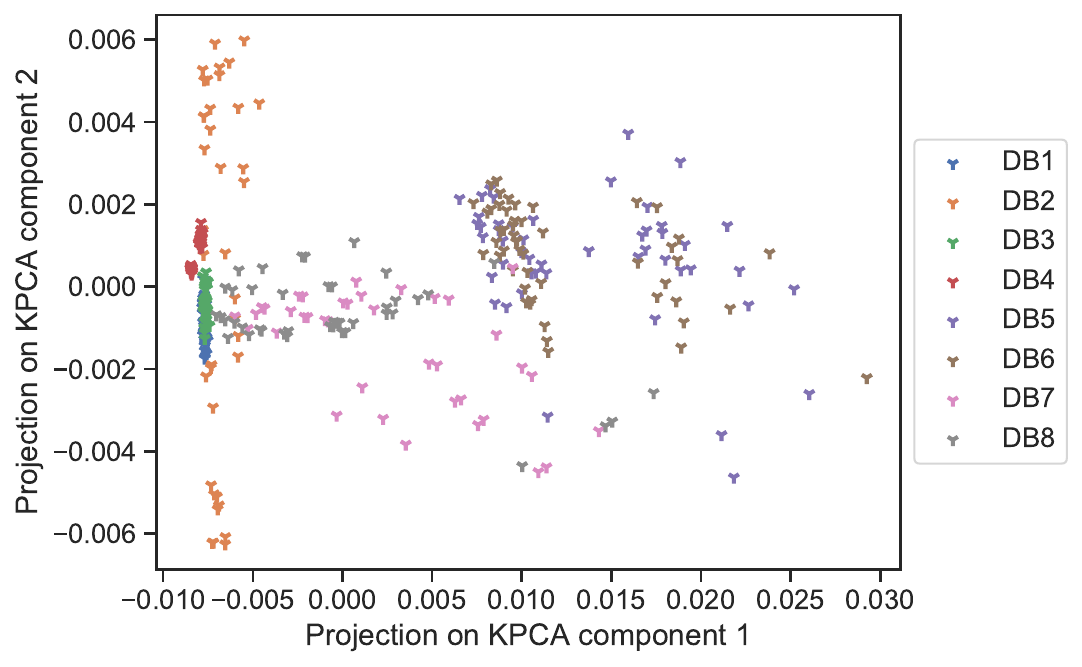}
    \caption{Kernel principal component analysis (KPCA) of training data in the 3-species SOAP descriptor space. This approach allows us to identify gaps in the \blue{training configuration space; {\it e.g.} DB7-8 were included to sample the gap} between the DB1-4 and DB5-6 clusters.}
    \label{fig:kpca}
\end{figure}

The first step to training a MLIP is to assemble a training database of high accuracy data relevant to properties of interest. Our training database comprises configurations relevant to the bulk, finite temperature and point defect properties of the alloy. It can be categorised into eight sub-databases (DB$x$) based on configuration type, as summarised in Table \ref{tab:database}. Fig.~\ref{fig:kpca} shows the kernel principal component analysis (KPCA) of the training database configurations using the SOAP kernel \cite{bartok_representing_2013}, coloured by subdatabases. The horizontal axis KPCA component can be interpreted as being roughly correlated with the temperature of the configurations, while the vertical axis KPCA component is correlated with strain. 

\begin{figure*}
    \includegraphics[width=0.99\linewidth]{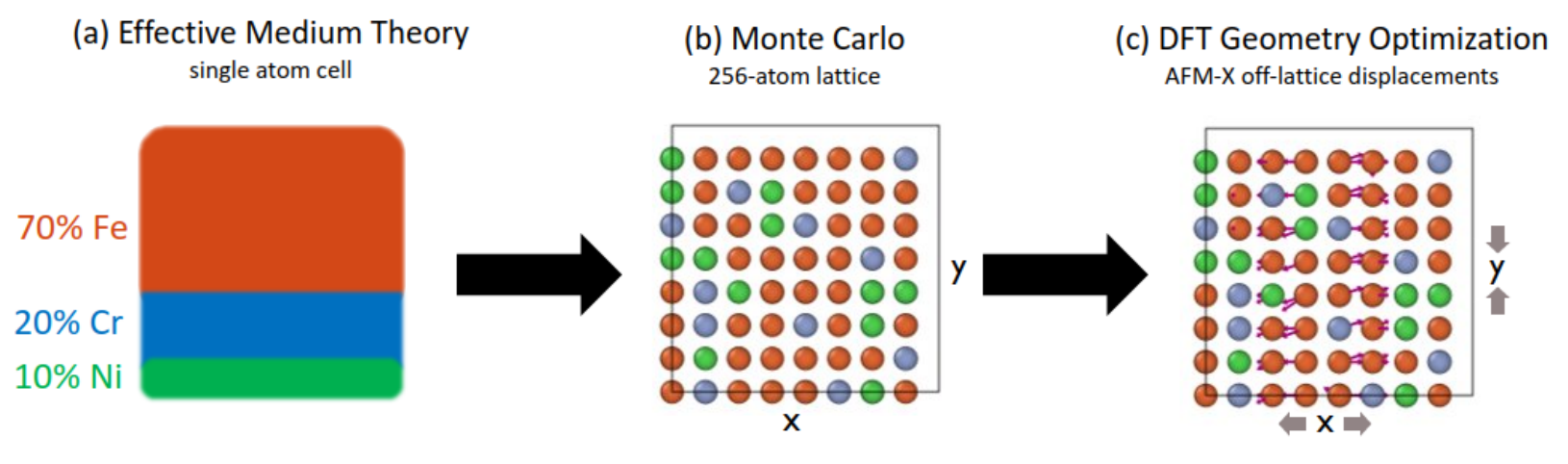}
    \caption{Protocol for generating bulk and vacancy training structures that sample different elemental configurations, with realistic short-range ordering derived from effective medium (EM) theory. In this case, the DFT relaxation off-lattice displacements shown by the red arrows of step 3 are for AFM layering along the $\hat{\mathbf{x}}$ axis (AFM-X).}
    \label{fig:em_protocol}
\end{figure*}

Perfect-lattice bulk and mono-vacancy supercells (containing 256 and 255 atoms respectively) with different element distributions that capture relevant short-range ordering were generated using an effective medium theory (EM) as described in subsection \ref{subsec:effective_medium}. These were fully relaxed as shown in the schematic of Fig.~\ref{fig:em_protocol}, using DFT calculations as described in subsection \ref{subsec:vasp}, to generate the training subdatabases DB1 and DB3 respectively. Random strains were applied to a subset of DB1 samples, to generate DB2, aimed at providing reference material which can inform an accurate description of the alloy's elastic behaviour. Snapshots from the DFT geometry optimisation trajectory of the EM monovacancy structures were compiled to form DB4, to provide details of the potential energy surface close to 0~K. 

The EM structures were generated to capture short-range ordering at temperatures of 600~K, 1000~K, 1500~K and 2000~K respectively. We ran ab-initio \blue{molecular dynamics (MD) for} the 1000~K-2000~K configurations from DB1 and DB3 at the temperatures they were generated for, and uncorrelated snapshots from these MD trajectories were compiled to form DB5 and DB6 respectively. In Fig.~\ref{fig:kpca}, we can see that the high temperature DB5-6 samples sit quite far apart from the low temperature DB1-4 samples along the first KPCA component. \blue{To aid the GAP model with better interpolation of training data, we need to populate such gaps between training data clusters along the principal axes of the kernel space. To this end, we extend the database by including snapshots from the geometry optimisation of a few bulk and monovacancy MD structures, comprising DB7 and DB8 respectively. These configurations are seen to sample the space between low and high temperature clusters in the KPCA plot Fig.~\ref{fig:kpca}, and their inclusion decreased the errors of our models.} The GAP potentials (both with and without the Fe spin extensions) were trained on this combined database DB1-8 as described in section \ref{subsec:gapfit}.

\subsection{\label{subsec:effective_medium}
Effective Medium Theory}

Past DFT studies on FeCrNi have modelled the alloy using special quasirandom structures (SQS)~\cite{piochaud_first-principles_2014,zhang_statistical_2021,manzoor_factors_2021}. A limitation of using SQS for magnetic materials is that in aiming to produce local environments reflective of the random alloy, they neglect atomic short-range order. It would be more effective to train a model using configurations in which temperature dependent short-range order is present, especially in the case of point defects where the energetics are correlated to the concentration of species decorating the defect~\cite{piochaud_first-principles_2014}. 

To obtain atomic configurations with physically motivated atomic short-range order, we draw samples from equilibrated, lattice-based Monte Carlo (MC) simulations, as depicted by the first step in Fig.~\ref{fig:em_protocol}.  A configuration is described by a set of lattice site occupancies, $\{ \xi_{i\alpha}\}$, where $\xi_{i\alpha}=1$ if site $i$ is occupied by an atom of chemical species $\alpha$, and 0 otherwise. The internal energy is described by a simple pairwise Bragg-Williams Hamiltonian~\cite{bragg_effect_1934, bragg_effect_1935},
\begin{equation}
    H = \sum_{i \alpha; j \alpha'} V_{i \alpha; j \alpha'} \xi_{i\alpha} \xi_{j \alpha'},
    \label{eq:bragg-williams}
\end{equation}
where the atom-atom interchange parameter, $V_{i \alpha; j \alpha'}$, describes the energy associated with an atom of species $i$ on site $\alpha$ interacting with an atom of species $j$ on site $\alpha'$. We assume these interchange parameters are isotropic, homogeneous, and have finite range, which simplifies Eq.~\ref{eq:bragg-williams}. 

The $V_{i \alpha; j \alpha'}$ are obtained using the $S^{(2)}$ theory for multicomponent alloys~\cite{khan_statistical_2016}, the details and implementation of which have been discussed extensively in earlier works~\cite{woodgate_compositional_2022, woodgate_short-range_2023, woodgate_interplay_2023, woodgate_integrated_nodate}. The theory uses the Korringa-Kohn-Rostoker (KKR) formulation of DFT, with disorder described via the coherent potential approximation (CPA)~\cite{faulkner_multiple_2018}, producing an effective medium representing the electronic structure of the disordered alloy. Our method captures the effects of an ordered magnetic state, in our case the single-layer antiferromagnetic (AFM) state, on the short range order state~\cite{woodgate_interplay_2023}. It is also possible to treat vacancies in this formalism at low concentrations by including a chemical species with no associated electrons, and no nuclear charge. 

In this work, we used the HUTSEPOT code~\cite{hoffmann_magnetic_2020} to construct the self-consistent potentials of DFT within the KKR-CPA formalism. The interchange parameters are fitted to the first four coordination shells, and they are tabulated in Appendix~\ref{appendix:interchange-parameters} for reference. The MC simulations used a cells of 256 atoms with periodic boundary conditions applied in all three directions. Simulations were equilibrated at the desired temperature and then decorrelated samples were drawn 25,600 MC steps apart, {\it i.e.} 100 MC steps per lattice site. The samples drawn from high temperature simulations will have little to no short-range order, while those drawn at lower temperatures will have a degree of short-range order present. \blue{These lattice-based EM configurations are then relaxed with DFT using setting described in the next section, to get the cell distortion and off-lattice atomic displacements as depicted in the last step of Fig.~\ref{fig:em_protocol}.} 

\subsection{\label{subsec:vasp}
DFT Settings}

The Vienna Ab-initio Software Package (VASP) \cite{kresse_ab_1994, kresse_efficiency_1996, kresse_efficient_1996} was used to compute ground state DFT structure, energy, atomic forces and cell stresses for the training database. \blue{MLIPs can only be as accurate as the data they are trained on, so the choice of DFT settings cap the achievable accuracy. A key choice when setting up DFT calculations is which exchange-correlation (XC) functional to use. For structural properties of transition metals, the Perdew–Burke–Ernzerhof \cite{perdew_generalized_1996} flavour of generalised gradient approximation (GGA) functionals is well established in the literature as a good choice, and so this is the XC functional we use. However, for more complex phenomena involving electronic excitations, we would require beyond-GGA approaches such as hybrid exchange, or even beyond-DFT methods such as the GW approach, and so we cannot hope to model such phenomena using a MLIP trained on GGA-based DFT.}

\blue{Depending on the range of atomic interactions that we hope to model, the choice of pseudopotential would also affect the DFT accuracy. For example, modeling nuclear cascades would require very short-range interactions that benefit from all-electron DFT codes or semi-core pseudopotentials \cite{olsson2016ab}. However, since we are targeting mechanical properties, the key players are the outer shell electrons of the alloying elements, so the accuracy we require is not compromised by using  softer pseudopotentials for the core electrons. We use} the standard VASP pseudopotentials based on the projector augmented wave (PAW) method \cite{kresse_ultrasoft_1999}, with eight, six and ten valence electrons for Fe, Cr and Ni respectively.

As Fe and Ni are magnetic materials, and Cr has shown a spin-density wave at low temperatures \cite{cottenier_what_2002, hafner_magnetic_2002}, it is essential to turn on the spin setting for more accurate DFT calculations of this alloy. Piochaud et al. reported that their tests with non-collinear spins generally relaxed to collinear  arrangements \cite{piochaud_first-principles_2014}, so we constrain our systems to collinear spin. Also, while austenitic steel is paramagnetic (PM) at operation temperatures of nuclear reactors, single-layer anti-ferromagnetic (AFM) ordering was found to be the most stable ground state \cite{piochaud_first-principles_2014}. To ensure we get ground state properties right, we impose an AFM ordering for DFT relaxations of our training samples. Initial magnetic moment magnitudes are overestimated to allow for relaxation, and are set at 3.0, 2.0 and 1.0 $\mu_{B}$ for Fe, Cr and Ni respectively. These atomic magnetic moments evolve during DFT electronic minimisation, and are computed in VASP by integrating the spin density within a sphere centred around the atom of interest. 

A Monkhorst-Pack $\mathbf{k}$-point mesh \cite{monkhorst_special_1976} of size $3\times3\times3$ was used to sample the Brillouin zone for all the training samples. Training data was computed with a planewave energy cutoff of 600~eV and an electronic self-consistency criterion of $10^{-7}$~eV. Preliminary steps, \textit{i.e.} initial geometry optimisations and ab-initio MD trajectories, were done using a lower self-consistency criterion of $10^{-4}$~eV. Electronic minimisation were performed using the preconditioned RMM-DIIS algorithm, while ionic position optimisations were done using the conjugate gradient method. The ab-initio MD trajectories were run for the NVT ensemble, using the Nose-Hoover thermostat \cite{nose1984unified,hoover1985canonical} and timesteps of either 2.5~fs (for T=1000~K,1500~K) or 5~fs (for T=2000~K). The initial $\sim200$ steps (exact number depending on when the total energy stabilises) of the trajectories were discarded, and uncorrelated samples were taken from the remaining 1000-1500 steps of each trajectory. These were then evaluated with the higher precision DFT settings before including in the training database.

\subsection{\label{subsec:gapfit}
Potential Fitting}

This section outlines how the GAP fits the Born-Oppenheimer potential energy surface (PES) by mapping the DFT training structures to their DFT energies, forces and stresses. As with other interatomic potentials, the GAP assumes that the energy $E_{N}$ of a configuration with $N$ atoms can be written as 
\begin{equation}
    E_{N} = \sum_{i=1}^{N} \epsilon(\mathbf{\hat{q}}_{i})
\label{eq:esum}
\end{equation}
where $\epsilon(\mathbf{\hat{q}}_{i})$ is the energy contribution from atom $i$ in the configuration, and $\mathbf{\hat{q}}_{i}$ is a descriptor vector that captures the unique features of the atom $i$ environment. The atomic energies are modelled as a linear combination of kernels \cite{rasmussen_gaussian_2006} given by
\begin{equation}
    \epsilon (\mathbf{\hat{q}}_{i})
    = \sum_{j}^{M} \alpha_{j} K(\mathbf{\hat{q}}_{j},\mathbf{\hat{q}}_{i})
    = \mathbf{k}_{i}^T \bm{\alpha}
\label{eq:edot}
\end{equation}
where the kernel basis functions $K(\mathbf{\hat{q}}_{j},\mathbf{\hat{q}}_{i})$ capture the similarity between environment $i$ and the $M$ environments of the basis set. As GAP training databases are generally large (eg. $N \sim 160,000$ atomic environments in this work), the usual strategy is to use only a small portion $M \ll N$ of the training set for the basis set. This makes the fit computationally tractable, while still retaining accuracy as the training database contains many groups of similar configurations that can be sparsified without much loss of information. These $M$ sparse points are selected by CUR decomposition, which searches for maximally dissimilar environments \cite{bartok_machine_2018}, ensuring that the chosen basis is representative of the variety across the full training database. 

Moving to the left-hand side of Eq.~\ref{eq:edot}: DFT does not give direct access to the atomic energies $\epsilon$, but rather, it gives us total energies $E_N$ of the training configurations, along with atomic forces and virial stresses. As forces and stress are properties that can be derived from atomic energies (\textit{e.g.} $\partial/\partial x$ for an atomic force along the Cartesian $x$ direction), we can back-calculate the unknown $N$ component vector $\bm{\epsilon}$ of atomic energies, from the known $D$ component vector $\mathbf{y}$ of DFT data, \blue{by the relation $\mathbf{y}=\mathbf{L}^T \bm{\epsilon}$, where $\mathbf{L}$  is a linear differential operator $\mathbf{L}$ }\cite{szlachta_accuracy_2014}. This allows us to derive the covariance kernel of the unknown target atomic energies ($\mathbf{K}_{NN}$) from the covariance kernel of the known DFT training data ($\mathbf{K}_{DD}$) using the relation \blue{$\mathbf{K}_{DD} = \mathbf{L}^T \mathbf{K}_{NN} \mathbf{L}$.} The sparse ridge-regression solution to Eq.~\ref{eq:edot} regularised by a diagonal tolerance matrix $\Lambda=\sigma_{\nu}^2 \mathbf{I}$ is
\begin{equation}
    \bm{\alpha} = \left[ \mathbf{K}_{MM} + 
    \mathbf{K}_{MN} \mathbf{L} \Lambda^{-1} \mathbf{L} \mathbf{K}_{NM}\right]^{-1}
    \mathbf{K}_{MN} \mathbf{L} \Lambda^{-1} \mathbf{y} .
\label{eq:alpha}
\end{equation}

With these optimised coefficients $\bm{\alpha}$, we can make predictions for new atomic environments $\mathbf{\hat{q}}_*$ using Eq.~\ref{eq:edot}. In this work, we use the Smooth Overlap of Atomic Positions (SOAP) descriptor 
\begin{equation}
    \mathbf{q}_{i} = \sum_{m=-l}^{l} (c_{nlm}^{i})^* c_{nlm}^{i} 
    \quad;\quad
    \mathbf{\hat{q}}_{i} = \frac{\mathbf{q}_{i}}{|\mathbf{q}_{i}|}
\label{eq:qsoap}
\end{equation}
where $c_{nlm}$ are expansion coefficients for the atomic neighbour density power spectrum \cite{bartok_representing_2013}. For computational tractability, this sum is truncated at radial-basis functions up to order $n<n_{\mathrm{max}}$ and spherical harmonics up to degree $l<l_{\mathrm{max}}$. The SOAP descriptor is suitable for describing atomic environments  since it is translationally, rotationally and permutationally invariant.  \cite{bartok_representing_2013}.
The corresponding covariance kernel is
\begin{equation}
    K(\mathbf{\hat{q}}_{i},\mathbf{\hat{q}}_{j}) = \delta^{2} 
    \left| \mathbf{\hat{q}}_{i} \cdot \mathbf{\hat{q}}_{j} \right|^{\zeta}
\label{eq:Ksoap}
\end{equation}
where $\delta$ is a factor that corresponds to the spread in energy of the training dataset, and $\zeta$ is tuned to amplify variations in the dot product, to pick out fine features.

While the SOAP kernel is good at capturing smoothly varying details of the PES, it is not as effective in capturing sharp features as well, such as exchange repulsion when atoms get close to each other, or phase transitions \cite{szlachta_accuracy_2014}. It has been found to be useful to use multiple kernels targeting different scales, so that each kernel does not have to simultaneously fit vastly different features \cite{bartok_machine_2018}. We hence use simple two-body kernels as well, to roughly capture the main trends of the PES, so that the SOAP kernels can focus on finer details. The two-body descriptor is just the pairwise separation between atoms, and we use the squared-exponential form
\begin{equation}
    K(\hat{q}_{i},\hat{q}_{j}) = 
    \delta^2 \exp\left({-\frac{1}{2}\left(\frac{\hat{q}_{i}-\hat{q}_{j}}{\theta_j}\right)^2}\right)
\label{eq:2body_gauss}
\end{equation}
for the two-body kernel, where $\theta_j$ is a lengthscale hyperparameter. For a 3-species alloy, we require 6 two-body kernels (one for each pair of species), and 3 SOAP kernels (one for each species). The kernel hyperparameters of Eq.~\ref{eq:qsoap}-\ref{eq:Ksoap} and Eq.~\ref{eq:2body_gauss} that we use are summarized in Table~\ref{tab:gap_params}. In addition these, we $\sigma_{\nu}^{\mathrm{energy}}=0.001$~eV/atom, $\sigma_{\nu}^{\mathrm{force}}=0.1$~eV/\AA, and $\sigma_{\nu}^{\mathrm{virial}}=0.05$~eV/atom for the regularization described in Eq.~\ref{eq:alpha}, values chosen based on the accuracy of our DFT training data. The MPI parallel \texttt{gap\_fit} code  from the QUIP software package \cite{klawohn2023gaussian, klawohn2023massively} was used to fit the GAP on the full dataset of Table~\ref{tab:database}.

\begin{table}
    \centering
    \caption{\label{tab:gap_params}Parameters for the collinear-spin GAP.}
    \begin{ruledtabular}
    \begin{tabular}{ccc}
     Parameter & Two-Body & Many-Body \\ \hline
     kernel choice & Gaussian & SOAP \\
     radial cutoff & 6.0 \AA & 6.0 \AA \\
     cutoff transition width & 1.0 \AA & 1.0 \AA \\
     $\delta$ & 1.0 eV & 0.1 eV \\
     sparse points & 100 & 800 \\
      $\theta$ & 1.0 & - \\
      $n_{\mathrm{max}},l_{\mathrm{max}}$ & - & 10,10 \\
      $\zeta$ & - & 4 \\
      $\sigma_{\mathrm{atom}}$ & - & 1.0 \AA\\  
    \end{tabular}
    \end{ruledtabular}
\end{table}

\begin{figure}
\includegraphics[width=\linewidth]{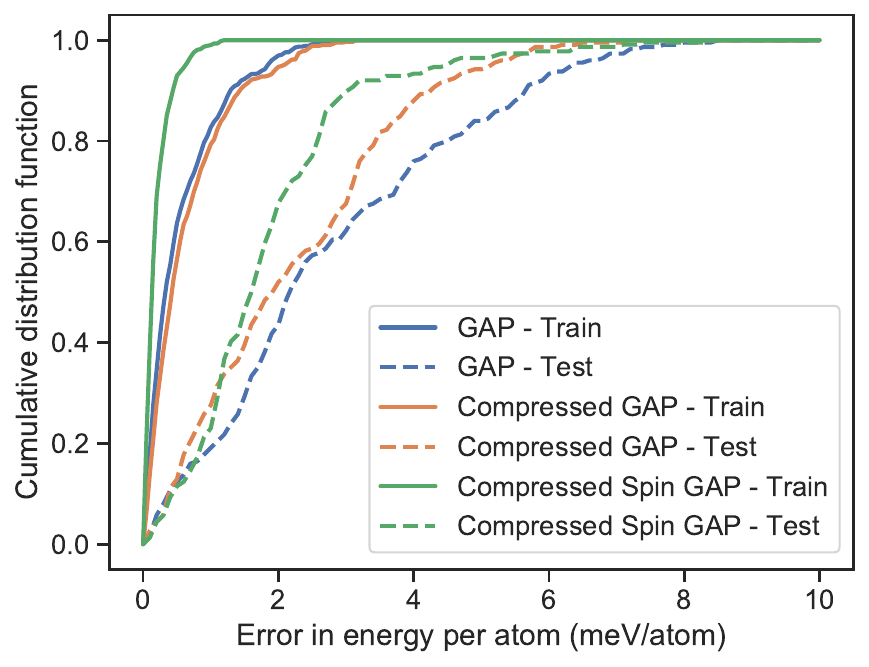}
\caption{\blue{Cumulative distribution functions of energy training (solid lines) and test (dashed lines) errors for the standard SOAP+GAP model (blue), a GAP model using compressed SOAP descriptors (orange) and a GAP model with an additional species included to differentiate between spin-up and spin-down Fe sites (Spin GAP; green). Compression does not significantly reduce accuracy (orange vs blue), and the Spin GAP model (green) performs significantly better.}}
\label{fig:train_test_cdf}
\end{figure}

The cumulative distribution functions of the \blue{training and test energy errors} of the resulting non-magnetic GAP are shown in Fig.~\ref{fig:train_test_cdf}. \blue{Similar plots for the force and stress errors have been included in Appendix \ref{appendix:errors}.} The GAP has energy training and test error within 2~meV/atom and 6~meV/atom respectively, which  are comparable to the differences in DFT energies due to choice of AFM layering direction. The train and test force errors are below 200~meV/\AA~for low temperature configurations and 400~meV/\AA~for high temperature configurations. The relatively large force errors for low-temperature configurations are because the GAP cannot fit the directional spin-induced face-centered tetragonal (FCT) ground state structure, as the GAP does not have spin degrees of freedom. The stress training and test errors are within 2~GPa and 4~GPa respectively, which are small as they are less than 2\% of the typical elastic constant values  that are about $\sim$200~GPa.

\subsection{\label{subsec:spin_model}
Spin Model}

As the SOAP and two-body descriptors do not have any information on spin, the standard GAP is spin-agnostic and cannot capture the tetrahedral deformation of AFM layering, and this limits the accuracy of its predictions. To address this, we introduce a fictitious fourth species to the GAP model, such that we have different descriptors for up spin (Fe$\uparrow$) and down spin (Fe$\downarrow$). This expanded descriptor set gives the model the flexibility to learn the tetrahedral deformation of AFM layering, and could be extended to the PM phase too. We henceforth refer to this 4-species potential as the Spin GAP. 

As the SOAP descriptor length scales quadratically with number of species, slowing down training and evaluation times prohibitively, we needed to use descriptor compression methods to make the training computationally tractable, such as those introduced in \cite{darby2023tensor}. We use combined radial and species compression with tensor product coupling across 300 mixed radial and species channels.  To check the effects of compression, a compressed version of the 3-species GAP was also fit. Fig.~\ref{fig:train_test_cdf} show that the compressed GAP has comparable errors to the standard GAP, verifying that there is no significant loss of accuracy when compression is used for this system. On the other hand, the Spin GAP, has significantly improved training and test errors over those of the standard GAP. Typical energy errors decrease by 1~meV/atom, force errors by about 50~meV/\AA, and stress training errors decrease by about 0.2~GPa.

\section{\label{sec:results}
results}

\subsection{\label{subsec:res_bulk}
Bulk properties}

\begin{table*}
    \centering
    \caption{\label{tab:bulk_prop}Lattice parameters and elastic constants of the FeCrNi alloy computed using DFT, GAP, Spin GAP and EAM-13 at 0~K. For comparison, DFT results for AFM Fe at 0~K and experimental values for finite temperatures have also been included. Dashes indicate that the value is equal to the previous value by symmetry.}
    \begin{ruledtabular}
    \begin{tabular}{ccccccc}
     & DFT-Fe & DFT & GAP & Spin GAP & EAM13 & Expt ($>$0 K)\\ \hline
     Phase & FCT & FCT & FCO & FCT & FCC & $\sim$FCC \footnotemark[1]  \\
     $a$ (\AA) & 3.42 & 3.50 & 3.49 & 3.50 & 3.55 & 3.52-3.59 \footnotemark[1] \\
     $b$ (\AA) & -- & -- & 3.53 & -- & -- & -- \\
     $c$ (\AA) & 3.68 & 3.58 & 3.56 & 3.58 & -- & -- \\
     $C_{11}, C_{22}, C_{33}$ (GPa) & 321, --, 254 & 259, --, 249 & 299, 243, 232 & 272, --, 268 & 333, --, -- & 204-226\footnotemark[2] \\
     $C_{12}, C_{13}, C_{23}$ (GPa) & 246, 96, -- & 164, 124, -- & 175, 163, 151 & 181, 126, -- & 196, --, -- & 132-134\footnotemark[2] \\
     $C_{44}, C_{55}, C_{66}$ (GPa) & 166, --, 262 & 147, --, 191 & 155, 171, 186 & 148, --, 188 & 161, --, -- & 111-122\footnotemark[2] \\
    \end{tabular}
    \end{ruledtabular}
    \footnotetext[1]{Temperature range 750~K-900~K. Ref. \cite{marucco_atomic_1994}.}
    \footnotetext[2]{Temperature range 4~K-295~K. Ref. \cite{ledbetter_stainlesssteel_1981,rajevac_lattice_2004}.}
\end{table*}

\begin{figure*}
\subfloat[\label{sfig:gopt_latt}]{%
  \includegraphics[width=0.49\linewidth]{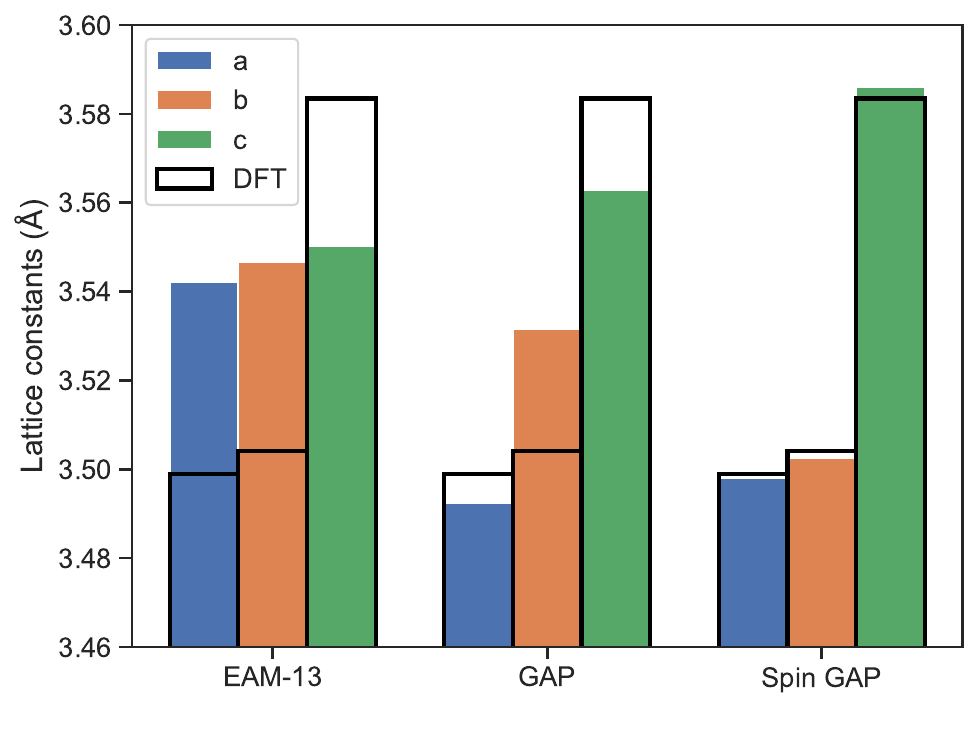}%
}
\hfill
\subfloat[\label{sfig:C}]{%
  \includegraphics[width=0.49\linewidth]{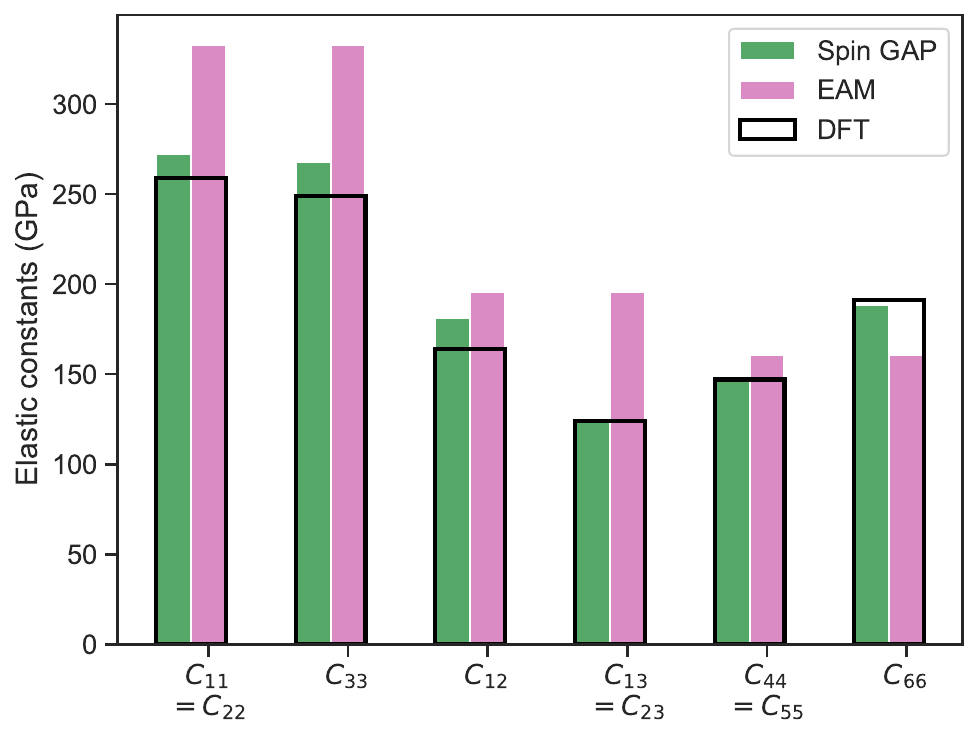}%
}
\caption{\blue{(a) Lattice constants (a,b,c) averaged over 400 configurations predicted by EAM, GAP and Spin GAP for the AFM state, with respect to the DFT AFM lattice constants averaged over 111 configurations. (b) The 6 independent elastic constants of FCT structures predicted by Spin GAP (green) and EAM (pink) from the average over 100 configurations each, compared to the DFT reference values computed for 1 configuration.}}
\end{figure*}

The ground state bulk properties of the FeCrNi alloy predicted by the GAP and the Spin GAP are summarised in Table \ref{tab:bulk_prop}, along with the corresponding DFT, EAM and experimental values for comparison, as well as DFT for pure Fe in an AFM spin state. The DFT results for the ground state structure of both AFM-Fe and AFM-FeCrNi are FCT, with the longer lattice constant corresponding to the AFM layering direction. This tetragonal distortion is more significant in pure AFM Fe ($\sim$ 0.26~\AA) compared to the alloy ($\sim$ 0.08~\AA), as the high concentration of Cr and Ni in the alloy lattice reduce the magnetism-induced geometry effects. Relaxation of EM structures with EAM-13 gives a lattice constant within the DFT and experimental range; however, EAM-13 fails to predict the lattice distortion, leading to a symmetric face centred cubic (FCC) lattice instead, \blue{as seen from the first set of bars in Fig.~\ref{sfig:gopt_latt}}. 

In comparison, the standard GAP \blue{predicts that a} distorted structure is more energetically favourable than a symmetric FCC structure, \blue{as it is trained on} relaxed DFT structures that have FCT lattices with zero atomic forces. However, since the standard GAP has no spin degrees of freedom, \blue{it does not have the framework to process the spin inputs required to elongate} one direction over the other two. Instead, the GAP \blue{training correlates} the distortion to minor differences in the alloy composition along the different axes in the relaxed training set geometries, and so predicts a face-centred orthohombic (FCO) relaxed lattice based on composition along the different axes of test structures. As seen \blue{from the second set of bars} in Fig.~\ref{sfig:gopt_latt}, the three lattice constants of the GAP relaxed FCO structure are equally spaced in a similar range to that of the two DFT relaxed FCT lattice constants. The Spin GAP on the other hand, has a more flexible model form that can capture the FCT distortion associated with alternating Fe$\uparrow$ and Fe$\downarrow$ layers. \blue{The third set of bars in Fig.~\ref{sfig:gopt_latt} show that} the Spin GAP successfully predicts lattice constants in excellent agreement with DFT, and the experimental values in Table \ref{tab:bulk_prop}.

The cumulative distribution functions of the non-zero elastic matrix components of EAM-13, Spin GAP and DFT are shown in \blue{Fig.~\ref{sfig:C}}. As the EAM predicts an FCC structure, it has only 3 distinct non-zero elastic constants \{$C_{11},C_{12},C_{44}$\}. The DFT and Spin GAP predict an FCT structure, and so have 6 distinct non-zero elastic constants \{$C_{11},C_{33},C_{12},C_{13},C_{44},C_{66}$\}. The standard GAP on the other hand has 9 distinct non-zero elastic constants as it predicts an FCO structure, the mean values of which are in Table \ref{tab:bulk_prop}.

As seen in \blue{the first two sets of bars} in Fig.~\ref{sfig:C}, the Spin GAP slightly overestimates $C_{11}$ and $C_{33}$ by $\sim$8\% each, but gets their relative magnitude right. This is a significant improvement from EAM-13 which overestimates the DFT $C_{11}$ by 30\%. The EAM-13 $C_{11}$ is in fact closer to that of pure AFM Fe reported in Table \ref{tab:bulk_prop}, indicating that EAM-13 does not capture the softening of this elastic mode due to alloying elements Cr and Ni, while the Spin GAP does capture this softening. Moving to the off-diagonal elastic components, the Spin GAP $C_{13}$ agrees very well with DFT, and is an improvement over the EAM-13 that overestimates it by 20\%. The Spin GAP performs less well with $C_{12}$, overestimating it by 10\%, but again this is a significant improvement over EAM-13 which overestimates this component by 58\%. Lastly, the Spin GAP $C_{44}$ and $C_{66}$ match very well with DFT \blue{as seen in the last two sets of bars in Fig. \ref{sfig:C}}, while EAM-13 predicts an averaged value between the DFT [$C_{44},C_{66}$] range. Overall, the GAP elastic constants agree well with DFT. The 8-10\% overestimation of modes \{$C_{11},C_{33},C_{13}$\} indicate that there are still errors in getting the behaviour of the long vs short axis of the spin-induced tetrahedral deformation. This is likely because our spin model is still quite a simple approximation to the full physics of the alloy --- it excludes contributions from Cr and Ni spins and ignores spin magnitudes. 

\begin{figure}

\subfloat[\label{sfig:ev}]{%
  \includegraphics[width=\linewidth]{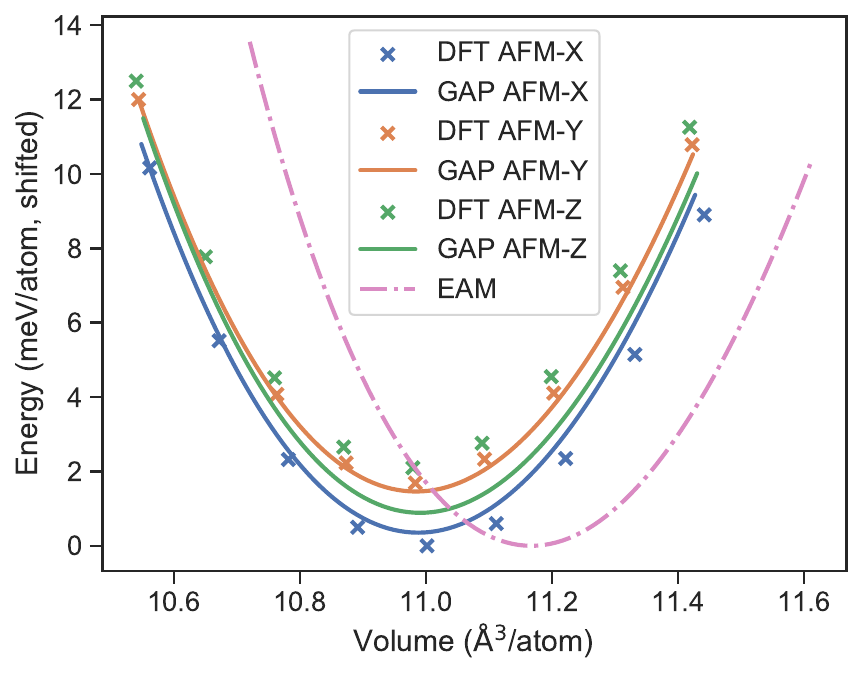}%
}
\vfill
\subfloat[\label{sfig:bulkmod}]{%
  \includegraphics[width=\linewidth]{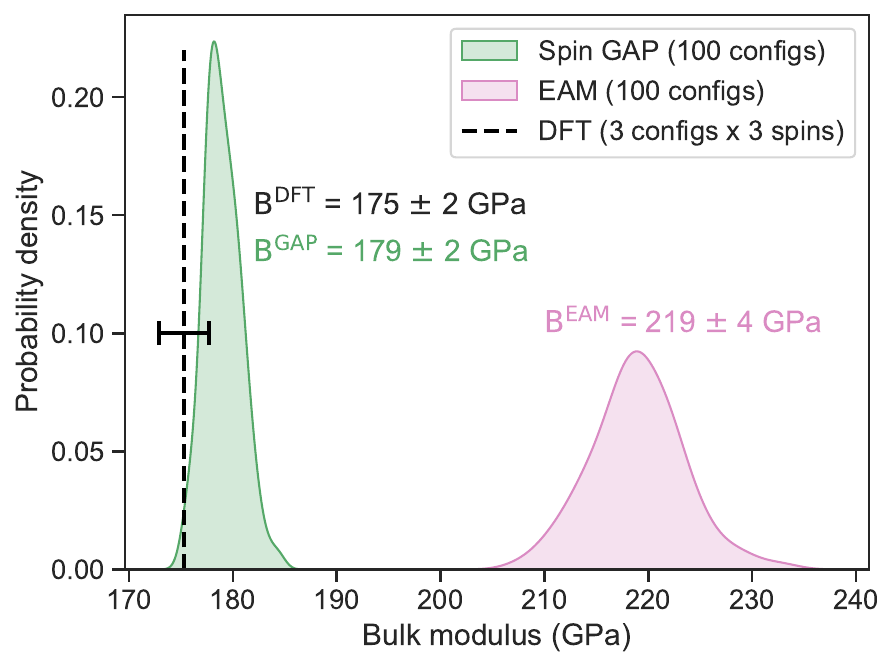}%
}

\caption{(a) Energy volume (E-V) curves predicted by DFT (crosses) and Spin GAP (solid lines) with AFM layering in $\hat{\mathbf{x}}$, $\hat{\mathbf{y}}$ and $\hat{\mathbf{z}}$ directions, and the corresponding spin-agnostic EAM-13 (dotted pink line). (b) Bulk modulus computed from a Birch-Murnaghan equation of state fit to the E-V curves.}
\label{fig:ev_curve}
\end{figure}

Given that the Spin GAP predicts the different elastic constants well, it unsurprisingly also predicts the equation of state reasonably well. Fig.~\ref{sfig:ev} compares the three Spin GAP E-V curves (AFM spin layering along $\hat{\mathbf{X}}$, $\hat{\mathbf{Y}}$ and $\hat{\mathbf{Z}}$ directions) of a 256-atom test configuration with respect to its DFT EV curves. We can see than the Spin GAP EV curves fall within the range of the DFT EV curves, even if they do not match exactly. The Spin GAP also gets the relaxed volume (E-V curve minimum) right, whereas the EAM-13 overestimates the relaxed volume as it is restricted to the FCC lattice. The bulk modulii computed from fitting an equation of state to these Spin GAP E-V curves are in very good agreement with DFT, overestimating it by just 4~GPa ($\sim$ 2\%), as seen in Fig.~\ref{sfig:bulkmod} (green distribution vs black dashed DFT reference), whereas the EAM-13 (pink distribution) overestimates the bulk modulus by 21\%. 

\begin{figure}
\includegraphics[width=\linewidth]{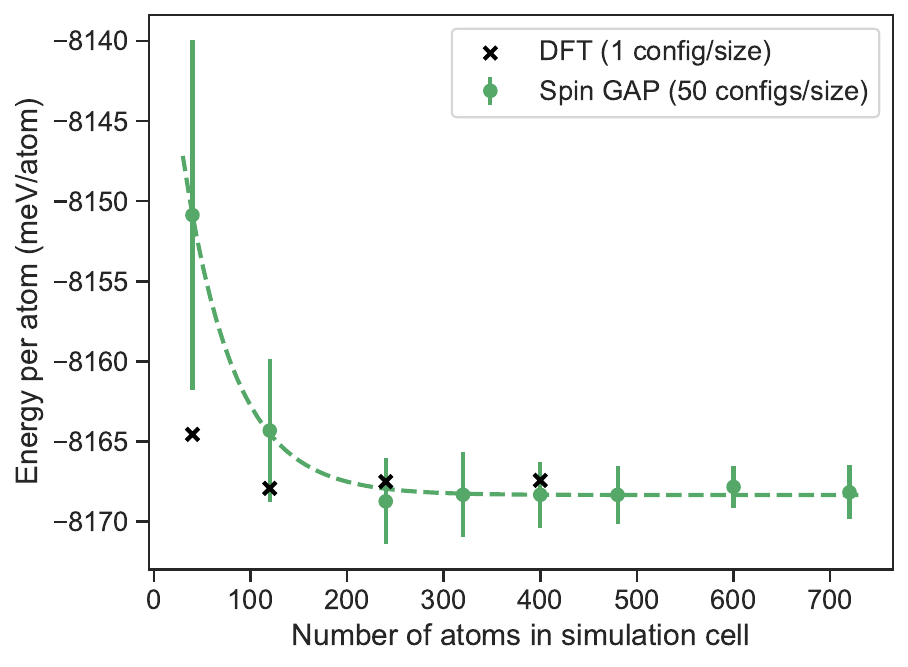}%
\caption{\blue{Convergence of Spin GAP ground state energy with respect to supercell size. The green dots and bars show the mean and standard deviation respectively, of the Spin GAP energy distribution of 40 configurations at each cell size. The Spin GAP energy converges to 1~meV/atom for cells with $>$200 atoms, and the converged energy agrees well with DFT (black crosses).} }
\label{fig:gopt_energy}
\end{figure}

As the Spin GAP is orders of magnitude faster than DFT, it allows for much more sampling of different random configurations, or larger supercells than the 256-atom ones used for DFT. Fig.~\ref{fig:gopt_energy} shows \blue{the mean and spread (one standard deviation) of the Spin GAP relaxed energy distributions at different supercell sizes, computed over 40 different configurations for each size. We see that the Spin GAP relaxed energy converges for supercells with more than 200 atoms, and its converged energy is in good agreement with the DFT reference values (black crosses) within the converged size range. For smaller sizes, as is the case when applying DFT to random alloys, finite size effects are significant leading to unconverged energy predictions. Also, the distributions at smaller sizes are broader, because each configuration's energy is computed over a smaller set of random local environments, leading to a larger variance between different configurations.}

\blue{For the cell size used for training (256-atoms), the full distribution of Spin GAP relaxed energies is in Fig.~\ref{sfig:E0_pm_kde}, and we can see that its peak agrees very well with the DFT mean.} The spin-agnostic GAP \blue{on the other hand} underestimates the ground state energy by about 2~meV/atom. We attribute this \blue{underestimation to the fact that, while the GAP correlates the tendency to distort to minor changes in composition, it associates FCT directionality to noise as it does not have spin degrees of freedom. The GAP training regularizes the forces of the low energy DFT training structures, and extrapolates to predicting FCO as a lower energy structure than FCT.} The Spin GAP overcomes this model form error by associating the tetragonal distortion with the AFM layering.

\begin{figure*}
\subfloat[\label{sfig:Esubs}]{%
  \includegraphics[width=0.49\linewidth]{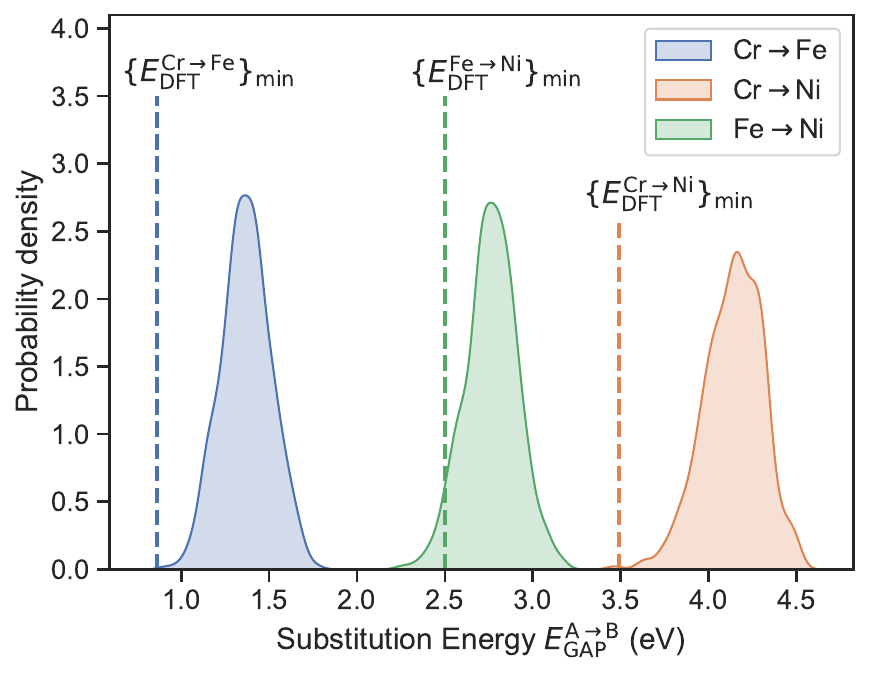}%
}
\hfill
\subfloat[\label{sfig:mu_violin}]{%
  \includegraphics[width=0.49\linewidth]{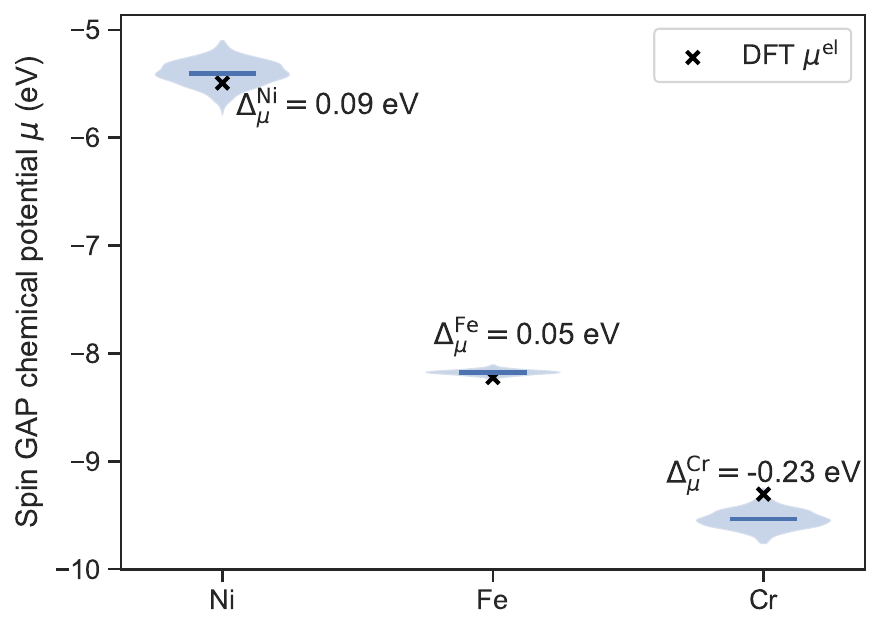}%
}
\vfill
\subfloat[\label{sfig:Ediff_violin}]{%
  \includegraphics[width=0.49\linewidth]{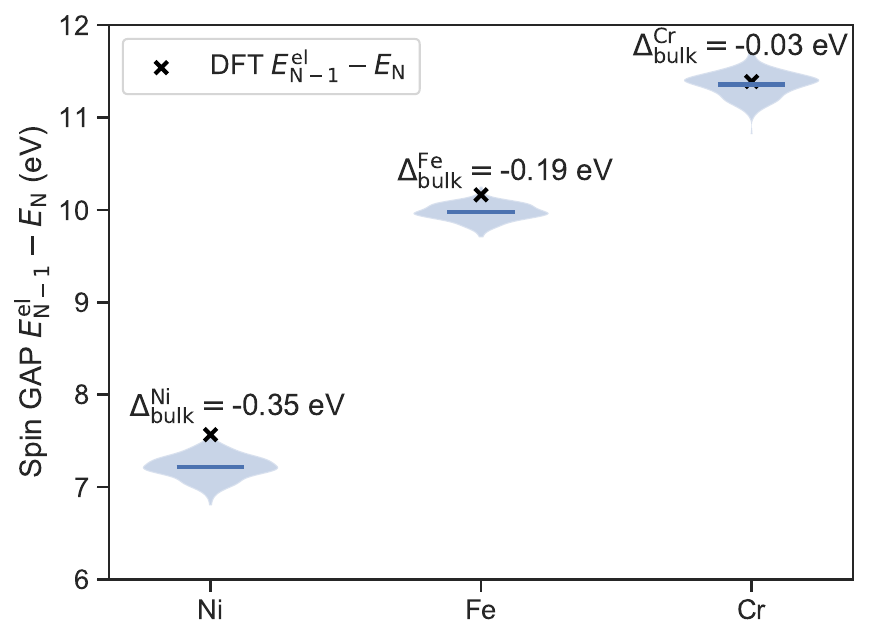}%
}
\hfill
\subfloat[\label{sfig:Evac_bars}]{%
  \includegraphics[width=0.49\linewidth]{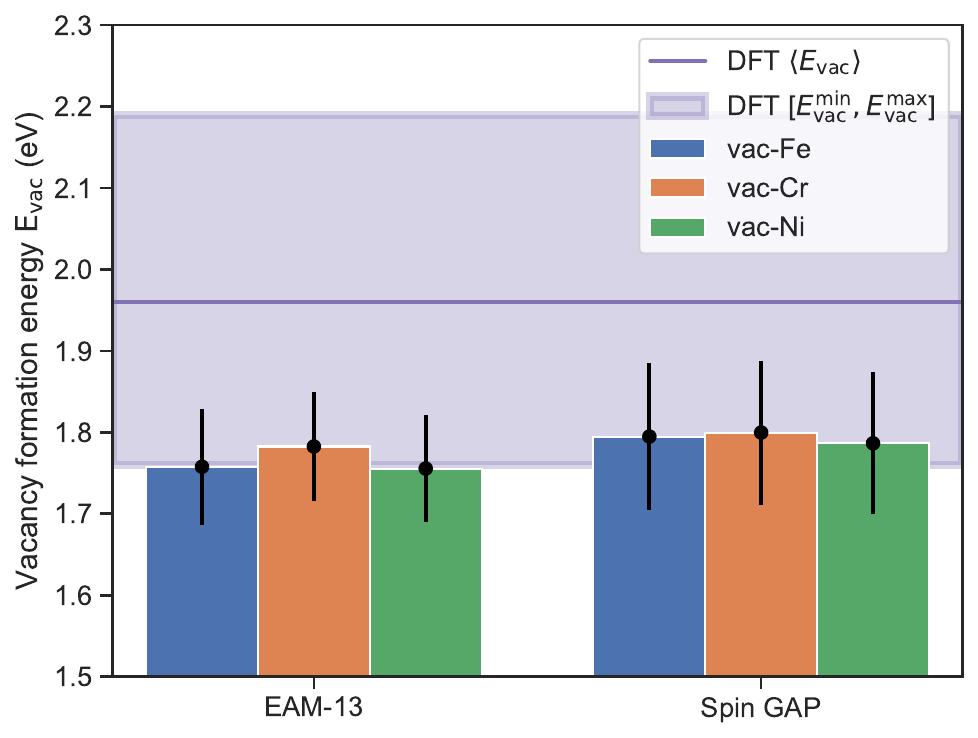}%
}

\caption{(a) Distribution of substitution energies for elemental swaps predicted by Spin GAP (100 configurations each) and corresponding DFT minima (dashed lines) computed from 40 DFT calculations performed by Piochaud el al. \cite{piochaud_first-principles_2014}. \blue{Distribution of (b) chemical potentials $\mu$ and (c) energy differences $E_{N-1}-E_N$ predicted by the Spin GAP from 800 configurations (256-atom) with respect to DFT means from 5 configurations (256-atom). (d) Mean and twice the standard deviations of Spin GAP and EAM-13 vacancy formation energies $E_\mathrm{vac}$ of 200 different EM configurations respectively, split by vacancy element type. The DFT mean and range of $E_\mathrm{vac}$ are taken from \cite{piochaud_first-principles_2014}.}}
\label{fig:chempot}
\end{figure*}

\subsection{\label{subsec:res_cpot}
Chemical potentials}

An essential ingredient for accurately computing energetics of point defects and their clusters is the chemical potential $\mu$ of each element in the alloy. This is the energy required to move an atom of the element from vacuum (infinitely far away from any interactions) to its bulk lattice site in the alloy supercell. The energy $E^{A\rightarrow B}$ required to substitute an element A with another element B \blue{at a lattice site in an alloy is equal to }
\begin{equation}
   \blue{ E^{\mathrm{A\rightarrow B}} = \mu^{\mathrm{B}} - \mu^{\mathrm{A}}. }
\label{eq:subs}
\end{equation}

\blue{This gives three equations for A,B $\in$ \{Fe,Cr,Ni\}, of which only two are independent.} To solve for the three unknown chemical potentials, we require three independent equations, and we can get this third equation from the sum of chemical potentials of the $N$ atoms in a configuration, which  gives the bulk energy $E_{N}$ of that configuration. For Fe$_{7}$Cr$_{2}$Ni, this is expressed by the equation 
\begin{equation}
    E_{N} = (0.7 \mu^{\mathrm{Fe}} + 0.2 \mu^{\mathrm{Cr}} + 0.1 \mu^{\mathrm{Ni}} ) N .
\label{eq:mu_sum}
\end{equation}

The chemical potentials $\mu^{\mathrm{Fe}}$, $\mu^{\mathrm{Cr}}$ and $\mu^{\mathrm{Ni}}$ can be determined for any lattice site in a cell by computing the two relevant substitution energies, and then solving the system of equations defined by Eq.~\ref{eq:subs}-\ref{eq:mu_sum}. For the Spin GAP, we solve this 3-species system of equations twice: first for \{Fe$\uparrow$,Cr,Ni\}, and then for \{Fe$\downarrow$,Cr,Ni\}, and we then take the average of the two resulting chemical potentials per species to get $\mu^{\mathrm{Fe}}$, $\mu^{\mathrm{Cr}}$ and $\mu^{\mathrm{Ni}}$. 

This procedure was carried out for randomly chosen lattice sites of 800 different configurations using the Spin GAP. The distributions of substitution energies are shown in Fig.~\ref{sfig:Esubs}, with reference DFT minimum values $E^{\mathrm{A\rightarrow B}}_{\mathrm{subs}}$ from \cite{piochaud_first-principles_2014}. We see that the minimum substitution energies from literature (using the same DFT settings as this work) agree well with the minima of the corresponding Spin GAP substitution energy distributions. The resulting chemical potential distributions are shown in the violin plots of Fig.~\ref{sfig:mu_violin}. DFT chemical potentials averaged over 5 configurations are also labelled using black crosses. We can see that the Spin GAP $\langle\mu^{\mathrm{Fe}}\rangle$ agrees very well with DFT. The Spin GAP $\langle\mu^{\mathrm{Ni}}\rangle$ and $\langle\mu^{\mathrm{Cr}}\rangle$ are slightly offset from the corresponding DFT averages by $+2$\% and $-2$\% respectively; this is probably a result of not accounting fully for the magnetic nature of Ni and Cr in this alloy.

\subsection{\label{subsec:res_vac}
Monovacancies}

To compute vacancy formation energies, we start with a relaxed bulk configuration of $N$ atoms and energy $E_N$, remove one atom of a given species `el', and then carry out a fixed-cell relaxation of the atomic positions to get the energy $E_{N-1}^{\mathrm{el}}$ of the vacancy configuration. As the vacancy configuration has one atom less than the corresponding bulk system, the comparison of the two configurations requires the chemical potential $\mu^{\mathrm{el}}$ of the atom removed to form the vacancy, giving the formula
\begin{equation}
    E_{\mathrm{vac}}^{\mathrm{el}} = \mu^{\mathrm{el}} + E_{N-1}^{\mathrm{el}} - E_N
\label{eq:vac}
\end{equation}
where quantities $(\mu^{\mathrm{el}} + E_{N-1}^{\mathrm{el}})$ and $E_N$ both have $N$ atoms each. We first compute the intermediate step $(E_{N-1}^{\mathrm{el}}-E_{N})$ for 800 different configurations using the Spin GAP and the results are shown in Fig.~\ref{sfig:Ediff_violin}. In this case, the Spin GAP result for Cr agrees well with DFT, whereas those for Fe and Ni are underestimated. These offsets are consistent with the Spin GAP underestimating $E^{\mathrm{Fe\rightarrow Ni}}$ slightly, as seen previously in Fig.~\ref{sfig:Esubs}. 

In fact, the small species-dependent errors in $\mu^{\mathrm{el}}$ and $E_{N-1}^{\mathrm{el}} - E_N$ cancel each other out to give a consistent behaviour in vacancy energy $E_{\mathrm{vac}}$, irrespective of which element was removed to form the vacancy. This is evident from the consistent mean and spread of the $E_{\mathrm{vac}}$ energy distributions denoted by the bars in Fig.~\ref{sfig:Evac_bars}. This cancellation of errors makes sense as $\mu^{\mathrm{el}}$ involves introducing an atom into the bulk whereas $E_{N-1}^{\mathrm{el}} - E_N$ involves the opposite procedure of removing an atom from the bulk structure. The resulting prediction of $E_{\mathrm{vac}}^{\mathrm{SpinGAP}}$=1.80~eV is in good agreement with the DFT value of $E_{\mathrm{vac}}^{\mathrm{DFT}}$=1.96~eV from the literature \cite{piochaud_first-principles_2014}. This prediction for $E_{\mathrm{vac}}$ is significantly more accurate than the standard spin-agnostic GAP trained on the same database which predicts $E_{\mathrm{vac}}^{\mathrm{GAP}}$=1.6~eV. The Spin GAP also performs marginally better than EAM-13, which gives a mean prediction of $E_{\mathrm{vac}}$=1.77~eV as seen in Fig.~\ref{sfig:Evac_bars}. We attribute the remaining error of -0.2~eV in the Spin GAP prediction to the incomplete magnetic model, as we do not account for spin polarisation of the electron density of Cr and Ni or for variations in Fe magnetic moment magnitudes near to defect sites. \blue{A DFT study on this alloy finds that the DFT vacancy formation energy drops from 1.98~eV down to 1.80~eV when they switch from imposing paramagnetic spins to ferrimagnetic spins on the same chemical composition, showing that changes in the magnetic state alone can indeed lead to an order of 0.2~eV changes in $E_{\mathrm{vac}}$ \cite{antillon2022b}, further suggesting that the Spin GAP's underestimation of $E_{\mathrm{vac}}$  is likely due to limitations of our spin model.}

\blue{On the other hand, a} recently reported fully non-collinear magnetic atomic cluster expansion (ACE) model for iron also makes similar magnitudes errors in vacancy formation energies (errors of 0.4~eV and 0.2~eV when trained without and with extended defect sub-databases respectively), despite being a much more rigorous magnetic model \cite{rinaldi2024non}. \blue{In their case, the errors probably come from challenges in sampling their multiplicatively larger spin space (as spin magnitudes and inter-spin angles are included). Whereas in our case, our simpler spin model comes with a relatively constrained spin space that is much easier to sample well. It is a promising result that we get comparable performance, suggesting that our spin model captures the main magnetic contributions despite being relatively simple. It would be interesting to try relaxing a few of the assumptions of our simple spin model while still keeping the spin space manageable, but that is beyond the scope of this paper and will be investigated in future works.}

\blue{\subsection{\label{subsec:res_sro}
Atomic Short Range Order}}

\begin{figure*}
\includegraphics[width=\linewidth]{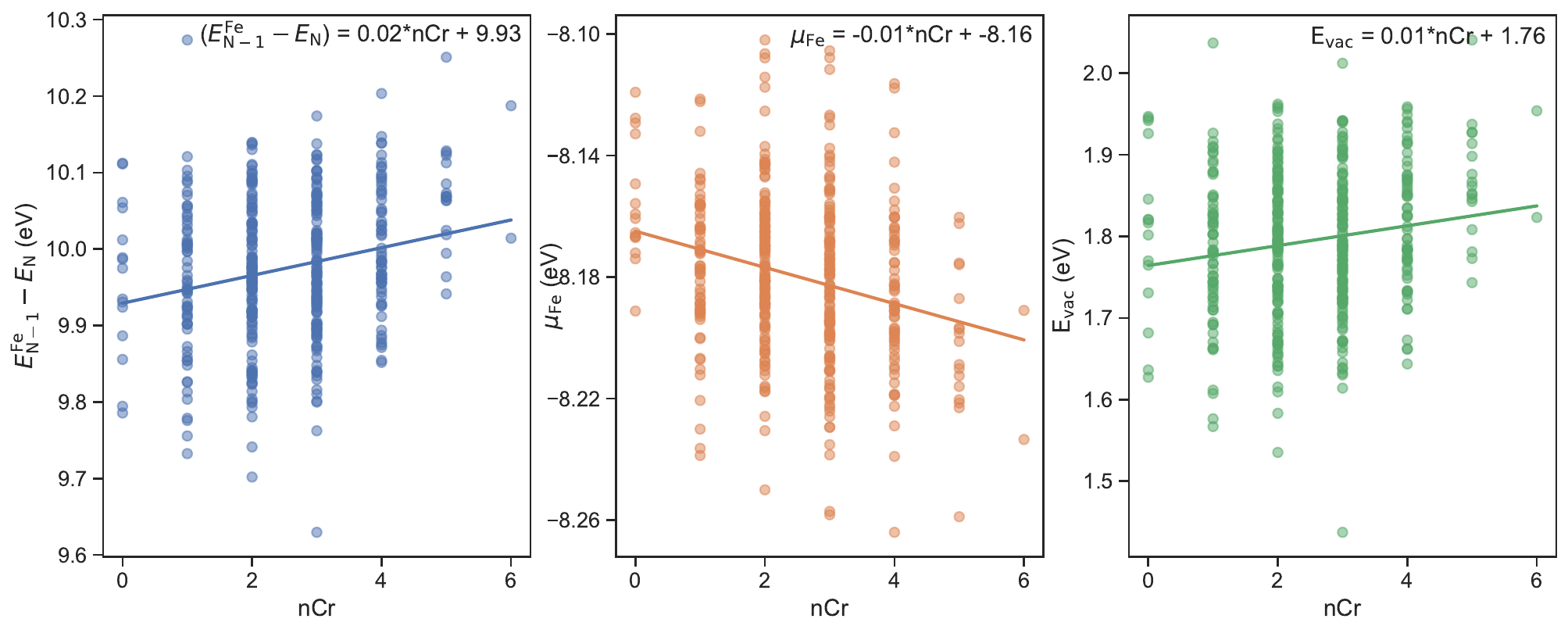}%
\caption{ \blue{Linear correlation between (a) energy difference of vacancy vs bulk configurations $ E_{N-1}^{\mathrm{Fe}}- E_{N}$ for vacancies generated by removing an iron atom each from the 200 different bulk configurations, (b) corresponding iron chemical potentials $\mu_{\mathrm{Fe}}$ for those vacancy sites, and (c) corresponding vacancy formation energy $E_{\mathrm{vac}}$ computed by adding the quantities of (a) and (b), all with respect to number of Cr atoms amongst the 12 nearest neighbours of the vacancy.} }
\label{fig:sro}
\end{figure*}

\begin{table}
    \centering
    \caption{\label{tab:sro_gradients} \blue{Linear correlation of vacancy formation energy with respect to number of first nearest neighbours (\#1nn) of type Cr and Ni respectively, from the literature \cite{piochaud_first-principles_2014,manzoor_factors_2021} (column 2-3) and our work (column 5), and for $(E_{N-1}^{\mathrm{Fe}} - E_N)$ from our work (column 4).}}
    \begin{ruledtabular}
    \begin{tabular}{ccccc}
     \blue{\#1nn}
     & \blue{$E_{\mathrm{vac}}$ (\cite{piochaud_first-principles_2014}) }
     & \blue{$E_{\mathrm{vac}}$ (\cite{manzoor_factors_2021}) }
     & \blue{$(E_{N-1} - E_N)$ }
     & \blue{$E_{\mathrm{vac}}$ } \\ \hline
     \blue{Cr} & \blue{0.05} & \blue{0.02} & \blue{0.02}  & \blue{0.01} \\
     \blue{Ni} & \blue{-0.04} & \blue{-0.04} & \blue{-0.01} & \blue{-0.004} \\
    \end{tabular}
    \end{ruledtabular}
\end{table}

\blue{DFT studies \cite{piochaud_first-principles_2014,manzoor_factors_2021} conclude that vacancies in the Fe$_{7}$Cr$_{2}$Ni are more stable in Ni-rich environments and less stable in Cr-rich environments. The gradients reported in the literature for the linear correlation between $E_{\mathrm{vac}}$ and number of Cr and Ni neighbours respectively have been included in Table \ref{tab:sro_gradients} for reference. The two studies agree well on the correlation with Ni neighbours, whereas Manzoor \textit{et.al.} find a weaker correlation with respect to Cr neighbours compared to Piochaud \textit{et.al.}, which they attribute to the fact that they have four times the amount of sampling \cite{manzoor_factors_2021}. In both cases however, they use constant pre-computed values for the chemical potentials, so trends in their vacancy formation energy Eq.~\ref{eq:vac} are effectively proportional to trends in just $E_{N-1}^{\mathrm{el}} - E_N$. }

\blue{The computational expense of calculating chemical potentials (two fixed-cell relaxations of element substitutions Eq.~\ref{eq:subs} for each vacancy) makes it impractical to compute it for every vacancy using DFT, but this is now possible to run cheaply with the Spin GAP. Fig.~\ref{fig:sro} from left to right shows the trends in energy difference $E_{N-1}^{\mathrm{el}} - E_N$, the corresponding chemical potentials, and the vacancy formation energies with respect to number of Cr neighbours as predicted by the Spin GAP. These results show that the chemical potential is also correlated to the number of Cr neighbours, with the opposite slope to the energy difference $E_{N-1}^{\mathrm{el}} - E_N$. Computing vacancy formation energy involves adding up these two quantities (Eq.~\ref{eq:vac}) and so their opposing slopes offset each other, reducing the overall magnitude of positive correlation between $E_{\mathrm{vac}}$ and number of Cr neighbours. Therefore, while we get consistent gradient of 0.02~eV/Cr for our $E_{N-1}^{\mathrm{el}} - E_N$ and the $E_{\mathrm{vac}}$ from the literature (see values in Table \ref{tab:sro_gradients}), the inclusion of variations in chemical potentials decreases the overall correlation. This makes sense as the chemical potential accounts for majority of the heterogeneity in the vacancies neighbourhood. The remnant positive correlation is likely because of the volume of the vacancy which has previously been seen to affect stability \cite{manzoor_factors_2021}: as Cr atoms are the smallest species in the alloy, they relax less into the vacancy, leading the larger vacancies than are less stable and have a higher formation energy.}

\blue{The inverse trends (but to a weaker extent) of the three subplots of Fig.~\ref{fig:sro} are seen with respect to number of Ni neighbours, the gradients of which are reported in Table~\ref{tab:sro_gradients}. The Spin GAP predicts a smaller negative correlation of -0.01~eV/Ni for $E_{N-1}^{\mathrm{el}} - E_N$ compared to the -0.04~eV/Ni values in the literature. Similar to the discrepancy between the two DFT studies for Cr trends, our lower correlation could be due to more variety in sampling. Piochaud \textit{et.al.} generate all their vacancy structure by removing different atoms from one 256-atom SQS structure, and Manzoor \textit{et.al.} do the same over two 256-atom SQS structures, leading to a large overlap in the neighbourhood of the vacancies they consider. On the other hand, our Spin GAP results were computed for vacancies introduced in 200 different starting structures from effective medium theory, and so sample a larger variety in long range vacancy environments. Despite the weaker correlations, we do still get a small negative relation of -0.004~eV/Ni for $E_{\mathrm{vac}}$ and number of Ni neighbours. Similar to the previous reasoning on stability of vacancy volumes, Ni atoms are the largest species in the alloy, so having more Ni neighbours leads to smaller vacancy volumes which are marginally more stable.} 

\blue{\subsection{\label{subsec:res_sro}
Alloy Composition}}

\begin{figure}
\includegraphics[width=\linewidth]{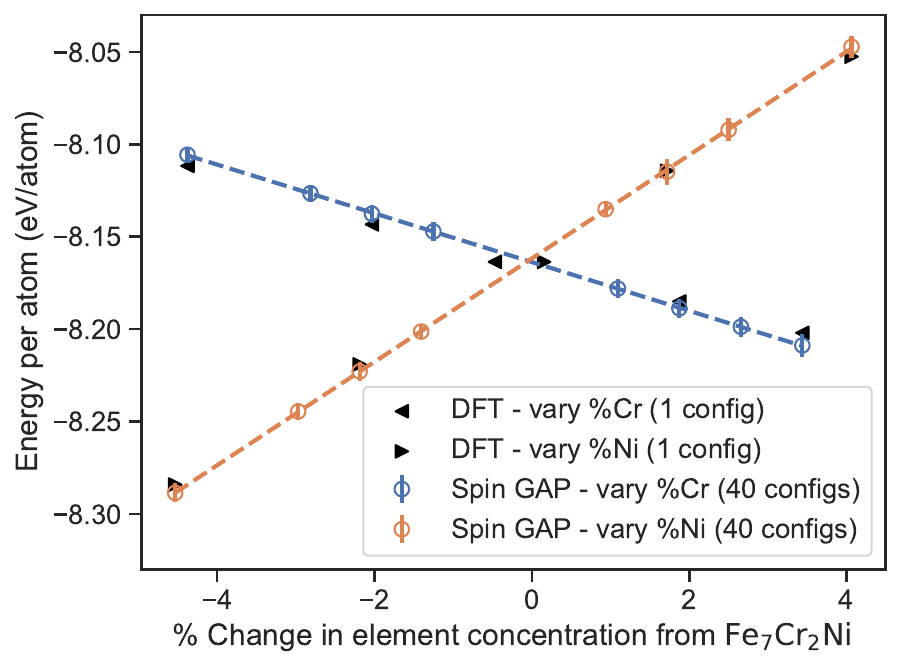}%
\caption{\blue{Variation in ground state energy per atom of bulk alloy with variation of Cr$\rightarrow$Fe concentrations (blue) and Ni$\rightarrow$Fe concentrations (orange), where 0\% refers to the target 70:20:10 ratio of Fe:Cr:Ni. We see good agreement with DFT values (black triangles) for this range of $\pm$4\% variation in composition.} }
\label{fig:vary_comp}
\end{figure}

\blue{In the 304 and 316 austenitic stainless steel grades, the typical Cr and Ni concentrations can vary upto $\pm$4\% from the model 70:20:10 ratio of Fe, Cr and Ni in our training set. It would hence be useful to check if our Spin GAP can extrapolate to small variations in the alloying ratios. Fig.~\ref{fig:vary_comp} shows the results for bulk relaxation for configurations within $\pm$4\% variations in  Cr (blue) and Ni (orange) concentrations. We see very good agreement with the DFT reference values in this range, validating that the Spin GAP can indeed be used to extrapolate to compositions about the training ratio. This is because the GAP framework breaks down each configuration into local atomic environments (LAEs) as in Eq.~\ref{eq:esum}, and the training set is likely to have a few Cr or Ni rich/poor LAEs since we sampled a large number of EM configurations, and these LAEs inform the potential for variations in alloy composition about the target ratio.} 

\blue{The gradients of the linear trends in Fig.~\ref{fig:vary_comp} are 0.0132~eV/\%Cr and 0.0280~eV/\%Ni. These convert to 1.32~eV and 2.80~eV for single atom Cr$\rightarrow$Fe and Ni$\rightarrow$Fe swaps respectively, which are consistent with the respective mean substitution energies of Fig.~\ref{sfig:Esubs}. This linear trend indicates that for small variations in alloying concentrations, the substitution energies do not change much. The chemical potential however would vary, since their computation involves the bulk energy as well via Eq.~\ref{eq:mu_sum}, and these bulk energies vary linearly as seen in Fig.~\ref{fig:vary_comp}. Since we have accurate bulk and substitution energies for the $\pm$4\% concentration range, we can expect all derivative properties computed in the previous results sections to follow through reliably.}

\subsection{\label{subsec:res_pm}
Paramagnetism}

\begin{figure}
\subfloat[\label{sfig:E0_pm_kde}]{%
  \includegraphics[width=\linewidth]{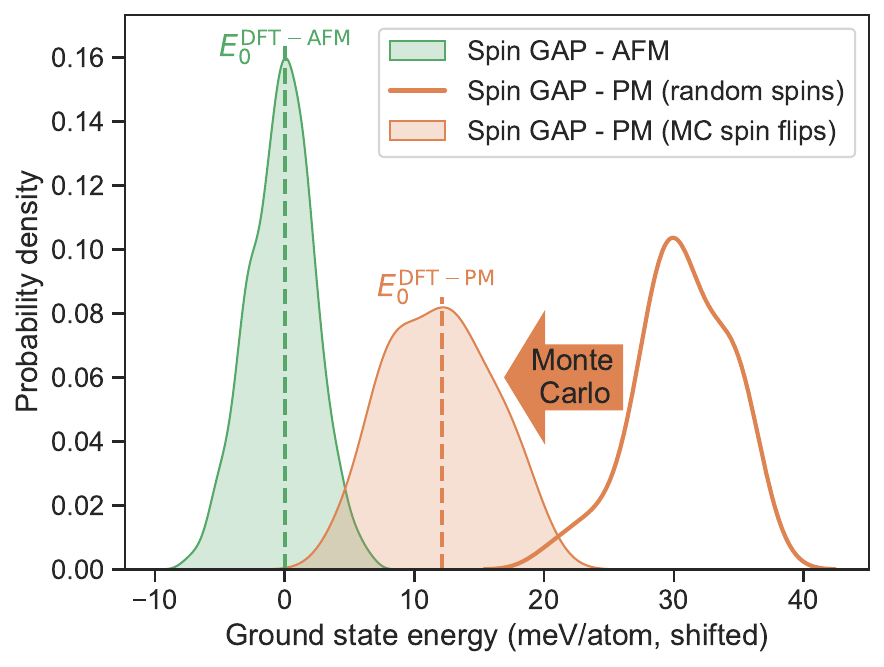}%
}
\vfill
\subfloat[\label{sfig:E0_pm_diag}]{%
  \includegraphics[width=\linewidth]{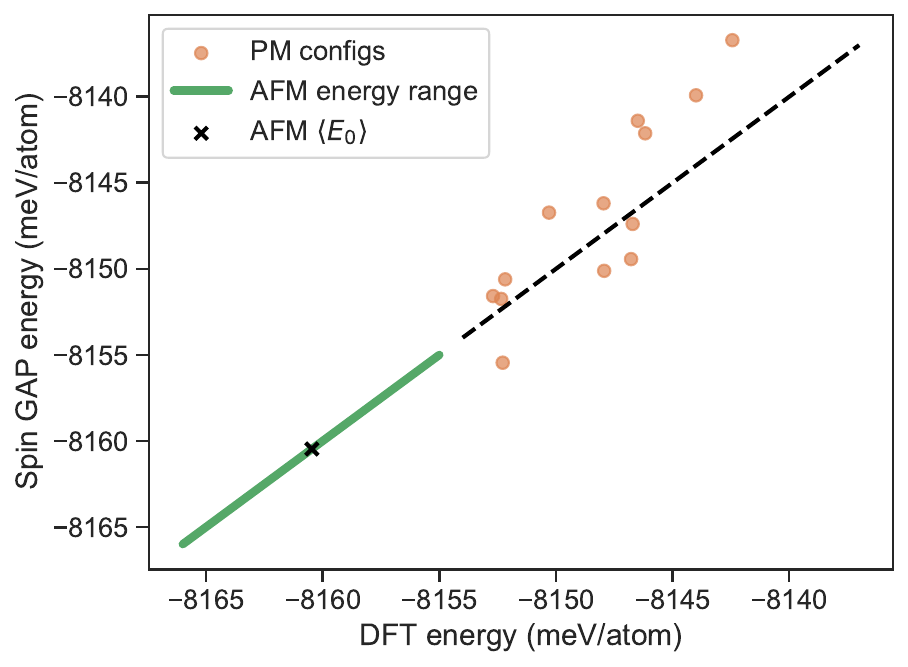}%
}
\caption{\blue{(a) DFT mean values (dashed lines) and Spin GAP distributions for ground state energies for AFM (green) vs PM states using uniform random 50\% up-down iron spins (unshaded orange) and after Monte Carlo optimisation with spin flip moves (shaded orange), comprising 400, 50 and 50 configurations respectively. (b) Correlation plot of Spin GAP relaxed energies for 13 PM reference DFT configurations with respect to their DFT relaxed energies.}}
\label{fig:para_ene}
\end{figure}

Although the Spin GAP was trained only on AFM DFT data, in this section we assess how well it extrapolates to the paramagnetic (PM) state. \blue{Comparing the magnetic ground state energies, Fig~\ref{sfig:E0_pm_kde} shows that} the Spin GAP correctly predicts the PM ground state of this alloy to be of higher energy, and hence less stable at 0K, than the AFM ground state. \blue{The unshaded orange distribution comprises energies of the starting configurations initialized with perfectly random 50\% up and down spin iron atoms, whereas the shaded orange distribution comprises their equilibrated energies after using Metropolis Monte Carlo with spin flip moves to optimise their paramagnetic spin states. We see that the equilibrated PM energy distribution agrees very well with the mean PM energy from DFT (orange dashed line in Fig.~\ref{sfig:E0_pm_kde}), showing that the Spin GAP can be used to generate reliable PM spin state configurations despite not being specifically trained for it. Fig.~\ref{sfig:E0_pm_diag} shows the direct correlation plot for energies of the 13 paramagnetic DFT test structures, where Spin GAP was used to evaluate these test structure with the same spin state as that from DFT. We see that the Spin GAP reproduces the DFT energies of these paramagnetic configurations very well, as all the points lie along the diagonal with the maximum deviation being 6~meV/atom.}

\begin{figure}
\includegraphics[width=\linewidth]{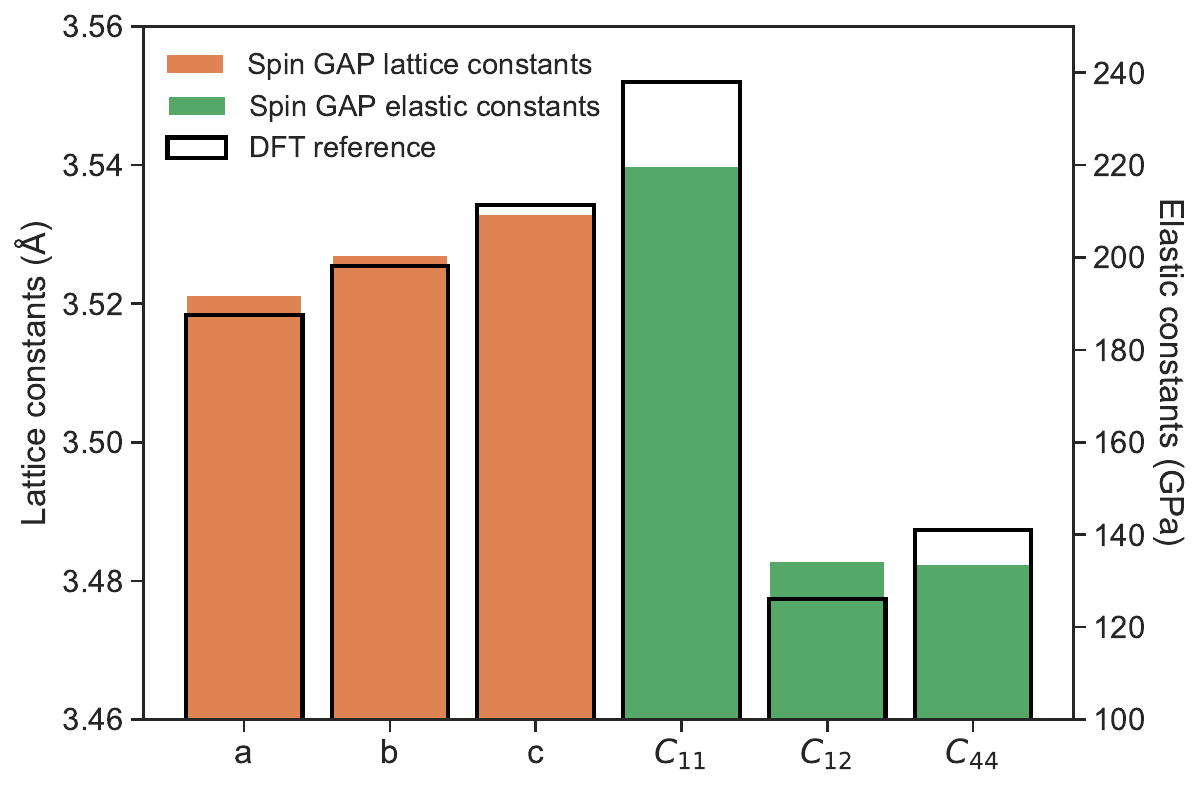}
\caption{\blue{Paramagnetic lattice constants (orange bars) and elastic constants (green bars) of the Spin GAP averaged over 50 configurations, compared to their respective DFT reference values (black bars). The DFT reference for lattice constants are an average over our 13 DFT PM test configurations, while those for the elastic constants are from the study by Antillon \textit{et.al.} \cite{antillon2022b}.}}
\label{fig:para_bar}
\end{figure}

\blue{For the PM ground state structure, DFT predicts a FCC structure with an average lattice constant of 3.53~\AA. The Spin GAP accurately predicts the paramagnetic FCC structure, with lattice constants that agree remarkably well with DFT, as seen from the orange bars of Fig.~\ref{fig:para_bar}. The lattice constants of all samples were sorted in ascending order and assigned to $a$, $b$ and $c$, respectively, and Fig.~\ref{fig:para_bar} shows their mean values. The small $\sim$0.005~\AA~difference between the three PM lattice constants is probably due to minor variations in ratio of the three species along the three axes, and this too is consistent between the Spin GAP and DFT. As seen from the green bars in Fig.~\ref{fig:para_bar}, the Spin GAP agrees reasonably well with the paramagnetic elastic constants as well, agreeing within 5-8\% to their respective DFT reference values from the study by Antillon \textit{et.al.} \cite{antillon2022b}. The fact that the Spin GAP reproduces these properties is a promising result, as the paramagnetic state is relevant to modelling the temperature range of most applications of Fe$_7$Cr$_2$Ni based austenitic steels.}

\section{\label{sec:conc}
Conclusion}

In summary, we have developed two machine learning interatomic potentials --- a standard GAP and a collinear Spin GAP --- for a model Fe$_7$Cr$_2$Ni alloy representative of austenitic steels. They are each trained on the same DFT database containing 159k atomic environments, and validated against as independent test set. We demonstrate the shortcomings of the standard GAP in not being able to capture the ground state tetragonal distortion due to AFM layering observed in this alloy, as the standard GAP formalism cannot account for magnetic spins. We then proposed an extended model incorporating spin to correct for this issue, and verified that the Spin GAP predicts the ground state structure, energy, elastic properties and vacancies in very good agreement with DFT. 

Atomistic modelling using this Spin GAP could be used to compute high accuracy inputs to larger scale microstructural models. \blue{Our Spin GAP is seen to extrapolate well to small variations in alloy composition making is useful for atomistic studies of more variety in austenitic steel grades. The spin model is also seen to describe the paramagnetic state well, which is relevant to modelling the alloy at operating temperatures of the steel components in nuclear and industrial machinery.} Beyond the properties validated in this paper, it could also \blue{be extended via iterative re-training to model more phenomena of interest such as} diffusion, phase transformations or responses to radiation. For instance, by \blue{adding and} testing the description of grain boundary structures, the Spin GAP could be used to study segregation of Cr at grain boundaries under radiation. Another avenue for future work would be to extend this Spin GAP to study hydrogen embrittlement in austenitic steel, relevant for applications in future fuel pipelines. This newly developed Spin GAP is a good starting point for efforts at modelling austenitic steel with close to DFT accuracy.

The DFT training data, interatomic potential and property predictions supporting this work are freely available from \url{https://doi.org/10.5281/zenodo.10577535}. For the purpose of open access, the author has applied a Creative Commons Attribution (CC-BY) licence to any Author Accepted Manuscript version arising from this submission.

\section{Acknowledgments}

L.S. and C.D.W. are supported by studentships within the UK Engineering and Physical Sciences Research Council-supported Centre for Doctoral Training in Modelling of Heterogeneous Systems, Grant No. EP/S022848/1.
J.R.K. acknowledges funding from the Leverhulme Trust under grant RPG-2017-191.
A.P.B. acknowledges support from the CASTEP-USER project, funded by the Engineering and Physical Sciences Research Council under the grant agreement EP/W030438/1.
L.S., C.S.B. and J.R.K. acknowledge funding from the ENTENTE consortium funded by the European Commission under grant agreement 900018.
L.S., C.B., C.D. and J.R.K. acknowledge PRACE support from IRENE at TGCC under Grant No 2020235585.
L.S. and J.R.K. acknowledge usage of the ARCHER2 facility for which access was obtained via the UKCP consortium and funded by EPSRC grant EP/X035891/1.
Calculations were performed using the Sulis Tier 2 HPC platform hosted by the Scientific Computing Research Technology Platform at the University of Warwick. Sulis is funded by EPSRC Grant EP/T022108/1 and the HPC Midlands+ consortium.
Further computing facilities were provided by the Scientific Computing Research Technology Platform of the University of Warwick.

J.R.K. designed the research, with input from all authors, and supervised L.S. (with support from A.P.B.). C.D.W. performed the KKR-CPA calculations and Monte Carlo simulations to generate the alloy supercells, with support from J.B.S. L.S. carried out the majority of the DFT calculations for the training database, fitted the GAP model (with support from A.P.B. and J.R.K.) and carried out the validation tests. C.D. carried out further DFT calculations, and together with C.S.B. supported on alloy and defect modelling.  L.S. drafted the manuscript with support from J.R.K. All authors revised the paper and approved its final version.

\appendix

\section{Effective Medium Theory Parameters}
\label{appendix:interchange-parameters}

Tabulated in Table~\ref{tab:interchange_parameters} are the pairwise atom-atom interchange parameters of Eq.~\ref{eq:bragg-williams} obtained for the Fe$_{7}$Cr$_{2}$Ni alloy considered in this work. The lattice-based model assumes interactions are isotropic and homogeneous, and we write $V^{(n)}_{\alpha\alpha'}$ to denote the energy associated with an atom of species $\alpha$ interacting with an atom of species $\alpha'$ at $n$th nearest neighbour distance. We find that it is sufficient to fit interactions to the first four neighbour distances to accurately capture the data produced using the $S^{(2)}$ theory.

\begin{table}
\centering
\begin{ruledtabular}
\begin{tabular}{lrrr}
$V^{(1)}_{\alpha\alpha'}$ & Fe & Cr & Ni \\ \hline
Fe                        & $  0.85$ & $ -5.28$ & $  4.58$ \\ 
Cr                        & $ -5.28$ & $ 23.44$ & $ -9.95$ \\
Ni                        & $  4.58$ & $ -9.95$ & $-12.18$ \\ 
                          &    &    &    \\
$V^{(2)}_{\alpha\alpha'}$ & Fe & Cr & Ni \\ \hline
Fe                        & $  0.12$ & $ -0.28$ & $ -0.29$ \\ 
Cr                        & $ -0.28$ & $  2.80$ & $ -3.64$ \\ 
Ni                        & $ -0.29$ & $ -3.64$ & $  9.33$ \\ 
                          &    &    &    \\
$V^{(3)}_{\alpha\alpha'}$ & Fe & Cr & Ni \\ \hline
Fe                        & $ -0.03$ & $ -0.16$ & $  0.53$ \\ 
Cr                        & $ -0.16$ & $  1.41$ & $ -1.70$ \\ 
Ni                        & $  0.53$ & $ -1.70$ & $ -0.30$ \\ 
                          &    &    &    \\
$V^{(4)}_{\alpha\alpha'}$ & Fe & Cr & Ni \\ \hline
Fe                        & $ -0.06$ & $  0.51$ & $ -0.59$ \\ 
Cr                        & $  0.51$ & $ -1.03$ & $ -1.50$ \\ 
Ni                        & $ -0.59$ & $ -1.50$ & $  7.15$ \\ 
\end{tabular}
\end{ruledtabular}
    \caption{Fitted pairwise atom-atom interchange parameters obtained from the effective medium theory for the composition Fe$_{7}$Cr$_{2}$Ni which are used in lattice-based Monte Carlo simulations to sample configuration space. All energies are in units of meV.}
    \label{tab:interchange_parameters}
\end{table}

\section{Potential Training \& Testing}
\label{appendix:errors}

\begin{figure}%
\subfloat[\label{sfig:Ferr_train}]{%
  \includegraphics[width=0.75\linewidth]{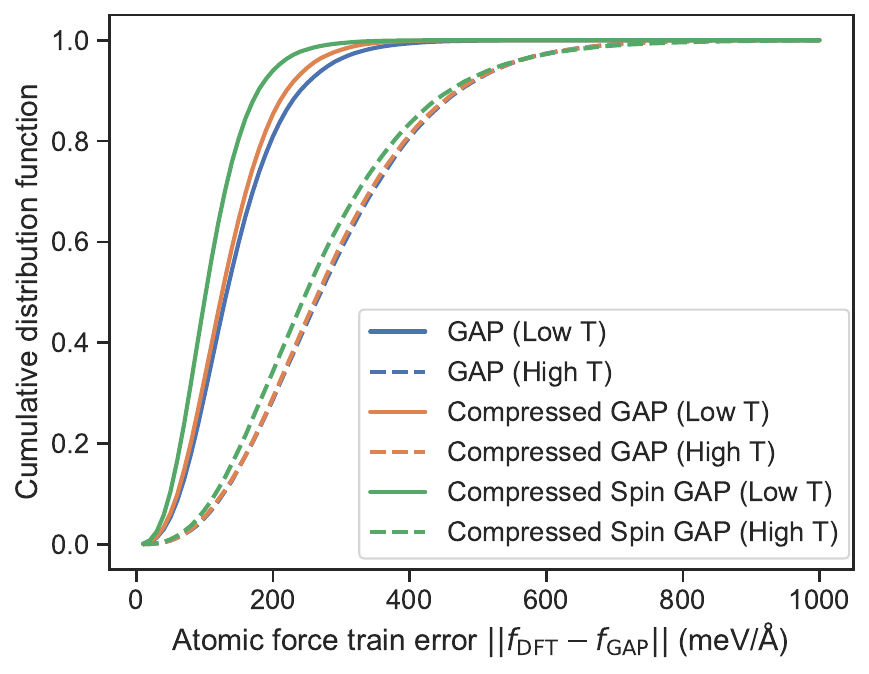}%
}
\vfill
\subfloat[\label{sfig:Ferr_test}]{%
  \includegraphics[width=0.75\linewidth]{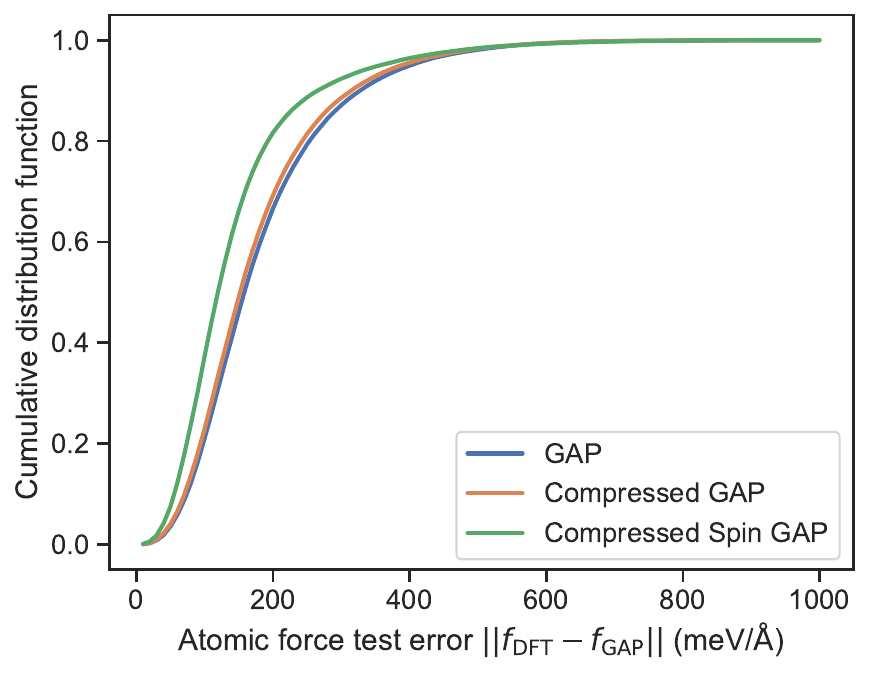}%
}
\vfill
\subfloat[\label{sfig:Serr_train}]{%
  \includegraphics[width=0.75\linewidth]{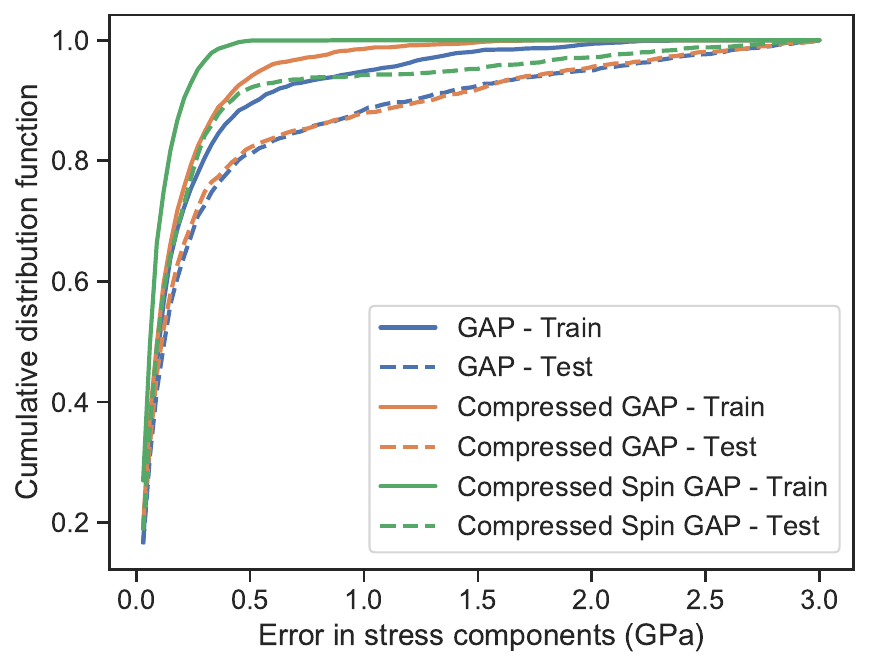}%
}
\caption{\blue{Cumulative distributions of (a) training and (b) test errors in forces, and in (c) stresses, for the standard SOAP GAP (blue), GAP with compressed SOAP descriptors (orange), and Spin GAP (green).}}
\label{fig:cdf_forces_stress}
\end{figure}

\blue{Fig.~\ref{fig:cdf_forces_stress} shows the training and testing errors for forces and stresses for the GAP, compressed GAP and Spin GAP. In all cases, the Spin GAP is seen to perform the best (lower errors). The errors at which these trends plateau were discussed previously in Section \ref{subsec:spin_model}. The composition of the database is reported in Table \ref{tab:database}.}

\newpage

%

\end{document}